\documentclass[a4paper,11pt]{article}

\usepackage{jheppub} 

\usepackage[T1]{fontenc} 

\usepackage{graphicx}
\usepackage{epsfig}
\usepackage{rotating}
\usepackage{amssymb}
\usepackage{subfigure}
\usepackage{dsfont}
\usepackage{psfrag}
\usepackage{amsmath,euscript,array,mathrsfs}
\usepackage{axodraw}
\usepackage{bbold}
\usepackage{epsf}

\newcommand{\beq}{\begin{equation}}
\newcommand{\eeq}{\end{equation}}
\newcommand{\beqs}{\begin{eqnarray}}
\newcommand{\eeqs}{\end{eqnarray}}
\newcommand{\lsim}{\mathrel{\raisebox{-
.6ex}{$\stackrel{\textstyle<}{\sim}$}}}
\newcommand{\gsim}{\mathrel{\raisebox{-
.6ex}{$\stackrel{\textstyle>}{\sim}$}}}
\newcommand{\Tr}{{\rm Tr}}

\def\hbar{\hspace{0pt}\raisebox{1pt}{$-$} \hspace{-7pt} h}

\def\r{\rho}

\newcommand{\be}{\begin{equation}}
\newcommand{\ee}{\end{equation}}

\newcommand{\bea}{\begin{eqnarray}}
\newcommand{\eea}{\end{eqnarray}}

\title{Spectrum-doubled Heavy Vector Bosons at the LHC}

\author[a]{Thomas Appelquist,}

\affiliation[a]{Department of Physics, Sloane Laboratory, Yale University, New Haven, Connecticut 06520, USA}

\author[b]{Yang Bai,}

\affiliation[b]{Department of Physics, University of Wisconsin-Madison, Madison, WI 53706, USA}

\author[a]{James Ingoldby,}

\author[c]{ Maurizio Piai\,}

\affiliation[c]{Department of Physics, College of Science, Swansea University,
Singleton Park, SA2 8PP, Swansea, Wales, UK}

\date{\today}

\abstract{
We study a simple effective field theory incorporating six heavy vector bosons together with the standard-model field content. The new particles preserve custodial symmetry as well as an approximate left-right parity symmetry. The enhanced symmetry of the model allows it to satisfy precision electroweak constraints and bounds from Higgs physics in a regime where all the couplings are perturbative and where the amount of fine-tuning is comparable to that in the standard model itself.

We find that the model could explain the recently observed excesses in di-boson processes at invariant mass close to 2 TeV from LHC Run 1 for a range of allowed parameter space. The masses of all the particles differ by no more than roughly $10\%$. In a portion of the allowed parameter space only one of the new particles has a production cross section large enough to be detectable with the energy and luminosity of Run 1, both via its decay to $WZ$ and to $Wh$, while the others have suppressed production rates. The model can be tested at the higher-energy and higher-luminosity run of the LHC even for an overall scale of the new particles higher than 3 TeV.
}

\begin{document}
\maketitle
\flushbottom

\newpage

\section{Introduction} 
\label{Sec:intro}

Recent analyses of $8$-TeV data by the ATLAS and CMS collaborations show 
a $2-3\sigma$
excess in the production of high-mass, $WW$, $WZ$, $ZZ$ as well as $Wh$ pairs at about 
$2$-TeV invariant mass~\cite{ATLAS,CMS}.
Inspired by these tentative signals of resonant production, we examine a 
simple effective field
theory (EFT) that combines the standard model (SM) with a minimal set of 
spin-one states at the
$2$-TeV scale. It describes the  coupling of these states to the 
longitudinal $W$ and $Z$ gauge bosons and 
the Higgs boson with a strength that can range from 
electroweak upward, leading
to total resonance widths ranging from tens of GeV to a few hundred 
GeV.  The coupling to the heavy states
is then limited to a perturbative range by precision measurements of 
standard-model parameters,
for example the Higgs-$W$-$W$ coupling. This restriction, in turn, insures 
that the fine tuning required to
keep the Higgs mass light is no worse than in the standard model itself.

EFTs of this type have been examined before~\cite{4site,LM,EFT}.  They can arise in various contexts, including 
little-Higgs theories~\cite{LH}, extra-dimensional theories~\cite{extradimensions}, 
and possibly walking technicolor theories~\cite{WTC} and theories of extended hidden local symmetry~\cite{HLS}.  Here, we adopt a minimalist approach: we focus on symmetric
setup in which six new vector bosons fill out a representation of
a new  $SU(2)_{L^{\prime}}\times SU(2)_{R^{\prime}}$ symmetry.
We impose on the EFT a parity symmetry broken only by the standard-model interactions 
and we introduce only a minimal set of new couplings
required to describe di-boson resonant behavior. This  naturally 
restricts the electroweak precision parameters to within current bounds.
Higgs boson decay widths remain in agreement with experiment provided that the coupling
strength of the new vector bosons is in the range  
 for which we can reasonably trust perturbation theory.

 We devote Sections \ref{Sec:EFT}--\ref{Sec:finetuning} to introducing the model and
 deriving and discussing analytical expressions for the physically relevant quantities.
 In Section~\ref{Sec:EFT}, we describe the EFT including its mass 
parameters and couplings. In Section~\ref{Sec:SM},
we describe the constraints on the couplings arising from the precision 
measurement of processes involving
standard-model particles. The decay widths of the new, heavy resonances 
are discussed in Section~\ref{Sec:decay},
and the fine tuning required to maintain the Higgs mass at $125$ GeV is 
described in Section~\ref{Sec:finetuning}.
  In Section~\ref{Sec:phenomenology} we perform a numerical study of the properties of the model.
  In particular we compute LHC production cross sections for the new resonances and show 
that the model could explain
excesses such as those possibly seen at LHC Run 1. 
We summarize and discuss our  findings in Section~\ref{Sec:discussion}.

\section{The Effective Field Theory}
\label{Sec:EFT}

\begin{table}
\begin{center}
\begin{tabular}{|c|c|c|c|c|c|c|}
\hline\hline
\multicolumn{7}{|r|}{{
\begin{picture}(0,55)(340,0)
\GCirc(70,26){10}{0}
\Line(80,26)(120,26)
\GCirc(130,26){10}{0}
\Line(140,26)(180,26)
\GCirc(190,26){10}{0}
\Line(200,26)(240,26)
\GCirc(250,26){10}{0.6}
\Text(100,38)[c]{ \Large $\Phi_L$}
\Text(160,38)[c]{ \Large $\Phi$}
\Text(220,38)[c]{ \Large $\Phi_R$}
\Text(70,6)[c]{ $SU(2)_L$}
\Text(130,6)[c]{ $SU(2)_{L^{\prime}}$}
\Text(190,6)[c]{ $SU(2)_{R^{\prime}}$}
\Text(70,46)[c]{ $SU(2)_L$}
\Text(130,46)[c]{ $SU(2)_{L^{\prime}}$}
\Text(190,46)[c]{ $SU(2)_{R^{\prime}}$}
\Text(250,46)[c]{ $SU(2)_R$}
\Text(250,6)[c]{ $U(1)_Y$}
\end{picture}
}}
\cr
\hline\hline
{\rm ~~~Fields~~~} &$SU(2)_L$ & $SU(2)_{L^{\prime}}$  & $SU(2)_{R^{\prime}}$ &  $SU(2)_R$ & $U(1)_{B-L}$ & $U(1)_{Y}$\cr
\hline\hline
$\Phi_L$ & $2$ & $2$ & $1$ & $1$ & $0$ & $0$ \cr
$\Phi$ & $1$ & $2$ & $2$ & $1$ & $0$ & $0$ \cr
$\Phi_R$ & $1$ & $1$ & $2$ & $2$ & $0$ & $\pm{1}/{2}$ \cr
\hline
$q_{L}$ & $2$ & $1$ & $1$ & $1$ & ${1}/{3}$ & ${1}/{6}$\cr
$q_{R}$ & $1$ & $1$ & $1$ & $2$ & ${1}/{3}$ & ${1}/{6}\pm{1}/{2}$\cr
$\ell_{L}$ & $2$ & $1$ & $1$ & $1$ & $-1$ & $-{1}/{2}$\cr
$\ell_{R}$ & $1$ & $1$ & $1$ & $2$ & $-1$ & $-{1}/{2}\pm{1}/{2}$\cr
\hline\hline
$W_{\mu}$ & $3$ & $1$ & $1$ & $1$ & $0$ & $0$\cr
$L_{\mu}$ & $1$ & $3$ & $1$ & $1$ & $0$ & $0$\cr
$R_{\mu}$ & $1$ & $1$ & $3$ & $1$ & $0$ & $0$\cr
$B_{\mu}$ & $1$ & $1$ & $1$ & $*$ & $0$ & $0$\cr
\hline\hline
\end{tabular}
\end{center}
\caption{Diagramatic representation and field content of the model. 
In the figure, the global symmetries are on the top, the gauge symmetries on the bottom.
The hypercharge is $Y=T^3+\frac{1}{2}(B-L)$, with $T^3$
the  generator of the $SU(2)_R$ group.
The fields $\Phi_L$, $\Phi$ and $\Phi_R$ are complex scalars, the quarks $q_i$ and leptons $\ell_i$ are 
Weyl spinors, while $W_{\mu}$, $L_{\mu}$, $R_{\mu}$ and $B_{\mu}$ are gauge bosons.
We complete the lepton doublet by adding right-handed neutrinos, which are singlets under all the gauge 
symmetries and hence inert. The $\ast$ highlights the fact that
 we gauge only the $U(1)_Y$ subgroup of $SU(2)_R\times U(1)_{B-L}$,
which implies that the individual $B_{\mu}$ gauge boson transforms as incomplete representation of $SU(2)_R$.
The presence of $B_{\mu}$ explicitly breaks the global $SU(2)_R$ symmetry. }
\label{Fig:4sites}
\end{table}

The ingredients of our EFT are listed in Table~\ref{Fig:4sites}. The columns and the
moose diagram at the top correspond to the global symmetry group $SU(2)_L\times SU(2)_{L^{\prime}}
\times SU(2)_{R^{\prime}}\times SU(2)_R\times U(1)_{B-L}$, 
in which the $SU(2)_L\times SU(2)_{L^{\prime}}\times SU(2)_{R^{\prime}} \times U(1)_Y$  subgroup is gauged. 
The hypercharge $Y=T^3+\frac{1}{2}(B-L)$ is a combination of the  $T^3$ generator of the $SU(2)_R$ group
and $U(1)_{B-L}$. We write the  gauge-boson
field $B_{\mu} \equiv B_{\mu}^3 [T^3+\frac{1}{2}(B-L)]$. 
The gauge fields of $SU(2)_L$ are $W_{\mu} \equiv W_{\mu}^{a} T^a$.
The additional vector fields  are $L_{\mu} \equiv L_{\mu}^{a}T^a$ and $R_{\mu} \equiv R_{\mu}^{a}T^a$.
The normalization is $\Tr \left[T^{a} T^{b} \right]= \delta^{ab}/2$. 

The Lagrangian for the bosons, including operators up to dimension 4, is
\beqs
{\cal L}_{b}&=&\label{Eq:Lag}
+2g\Tr W^{\mu}J_{L\,\mu}\,+\,2g^{\prime}\Tr B^{\mu} J_{Y\,\mu}\\
&&-\,\frac{1}{2}\Tr {W}_{\mu\nu}{W}^{\mu\nu}
\,-\,\frac{1}{2}\Tr {L}_{\mu\nu}{L}^{\mu\nu}\,
\,-\,\frac{1}{2}\Tr {R}_{\mu\nu}{R}^{\mu\nu}
\,-\,\frac{1}{2}\Tr {B}_{\mu\nu}{B}^{\mu\nu}\nonumber\\
&&+\,
 \frac{1}{4}\Tr |D\Phi_L|^2+ \frac{1}{4}\Tr |D\Phi|^2+ \frac{1}{4}\Tr |D\Phi_R|^2\,-\,V(\Phi_i)\,.\nonumber
\eeqs
The field strength tensors are defined so that the gauge bosons are canonically normalized and 
we denote with $g$, $g_{L}$, $g_R$ and $g^{\prime}$ the four gauge couplings.
 $J_{L\mu}$ and $J_{Y\mu}$ are electroweak matter currents bilinear
in the SM fermion fields.  Their mass terms will be discussed later. 
The covariant derivatives for the scalars are
\beqs
D_{\mu}\Phi_L &\equiv& \partial_{\mu}\ \Phi_L\,-\,i\left(g\, {W}_{\mu}\ \Phi_L - g_{L}\,\Phi_L{L}_{\mu}\right)\,,\nonumber\\
D_{\mu}\Phi &\equiv& \partial_{\mu}\ \Phi\,-\,i\left(g_{L}\,{L}_{\mu}\ \Phi - g_{R}\,\Phi{R}_{\mu}\right)\,,\\
D_{\mu}\Phi_R &\equiv& \partial_{\mu}\ \Phi_R\,-\,i\left(g_{R}\,{R}_{\mu}\ \Phi_R - g^{\prime}\,\Phi_R{B}_{\mu}\right)\,.\nonumber
\eeqs

We write the potential for the scalars in the following way:
\beqs
V&=&\frac{\lambda_L}{16} \left(\frac{}{}\Tr\left[\frac{}{}\Phi_L\Phi_L^{\dagger}-F_L^2\,\mathbb{I}_2\right]\right)^2
\,+\,\frac{\lambda}{16}  \left(\frac{}{}\Tr\left[\frac{}{}\Phi\Phi^{\dagger}-f^2\,\mathbb{I}_2\right]\right)^2\nonumber\\
&&\,+\,\frac{\lambda_R}{16}  \left(\frac{}{}\Tr\left[\frac{}{}\Phi_R\Phi_R^{\dagger}-F_R^2\,\mathbb{I}_2\right]\right)^2\,,
\eeqs
which means we do not include mixing terms in the potential between the scalars. These will be loop generated, and potentially require fine tuning, but in the coupling-strength range allowed phenomenologically this will be less than the amount of fine tuning we discuss in Section~\ref{Sec:finetuning}.

Taking $\lambda_L = \lambda_R$, $F_L=F_R\equiv F$ and $g_{L}=g_R=g_{V}$,
 the EFT is left-right symmetric, except for the weak gauging $g$ and $g^{\prime}$ of the SM groups $SU(2)_L$  and $U(1)_Y$.  
 The coupling strength $g_{V}$ is a free parameter 
that we will allow to range from electroweak strength to ${\cal O}(4 \pi)$. 
 The potential induces symmetry breaking
 at the scale $F$ in the case of $\Phi_L$ and $\Phi_R$,
 at the scale $f$  in the case of $\Phi$, and consequently generates the electroweak scale $v_{W} \simeq 246$ GeV.

The bifundamental $\Phi$ describes the Nambu-Goldstone bosons (NGBs) 
associated with the scale $f$ along with the physical Higgs particle $h$.  The bifundamentals $\Phi_L$ and $\Phi_R$ play a similar
role with respect to the scale $F$. Altogether the NGBs provide longitudinal components for all the gauge fields except for the massless photon. 
For simplicity, we take $\lambda_L = \lambda_R \rightarrow \infty$, freezing out the corresponding physical scalars and 
imposing the nonlinear constraints $\Phi_L\Phi_L^{\dagger}=F^2\,\mathbb{I}_2=\Phi_R\Phi_R^{\dagger}$.
In unitary gauge we replace $\Phi_L=F\,\mathbb{I}_2=\Phi_R$ and $\Phi=(f+h)\,\mathbb{I}_2$, 
with $h$ the (real and canonically normalized) Higgs field.

The $4\times 4$ mass matrix for the neutral vector bosons in the basis $(W^{\mu}, L^{\mu}, R^{\mu}, B^{\mu})$ is 
\beqs
{\cal M}^2_0&=&
\frac{1}{4}
\left(\begin{array}{cccc}
g^2 F^2& -g g_{V} F^2 & 0&0\cr
-g g_{V} F^2 & g_{V}^2 (f^2+F^2)& - g_{V}^2 f^2&0\cr
0&- g_{V}^2 f^2 & g_{V}^2 (f^2+F^2)&-g^{\prime}g_V F^2\cr
0&0&-g^{\prime}g_V F^2 & g^{\prime\,2}F^2
\end{array}\right)\,,
\eeqs
while the $3\times 3$ mass matrix ${\cal M}^2_+$ for the charged vectors can be obtained by removing the last row and column.

The eigenvalue structure of the $3\times 3$ charged-vector mass matrix depends on the two ratios $g^2/g_V^2$ and $f^2/F^2$,
where $g \approx 0.6$. The lightest eigenvalue is fixed to be
$M_W = 80.4$ GeV and the next to lightest eigenvalue  $M_{V_1}$ is taken to be at least of order  $2000$ GeV. 
In this range, at least one of the two ratios
$g^2/g_V^2$ or $f^2/F^2$ must be  small.

We next exhibit explicitly the mass eigenvalues of the charged sector.
With some abuse of notation,
we denote the mass
eigenstate $W^+$ with the same symbol as the original interaction eigenstate, 
though it results from mixing with the heavy vectors. 
In the regime $g\ll g_{V}$ the eigenvalues are  given by the following relations: 
\beqs
M_{{W}^+}^2&\simeq&\frac{1}{4}\,g^2\,\frac{f^2F^2}{F^2+2f^2}\,,~~~~
M_{V_1^+}^2\,\simeq\,\frac{1}{8}(2g_V^2+g^2)F^2\,,~~~~\nonumber\\
&&
M_{V_2^+}^2\,\simeq\,\frac{1}{4}g_{V}^2(F^2+2f^2)\,+\,\frac{1}{8}g^2\frac{F^4}{F^2+2f^2}\,,
\eeqs
and the heavy eigenstates for all charge assignments are given  
by $V_1 \simeq (L+R)/\sqrt{2}$, and   $V_2 \simeq (L-R)/\sqrt{2}$. 
In the regime $f\ll F$, we find
\beqs
M_{{W}^+}^2&\simeq&\frac{1}{4}\,\frac{g^2g_V^2}{g^2+g_V^2}\,f^2\,,~~~
M_{V_1^+}^2\,\simeq\,\frac{1}{4} g_V^2(F^2+f^2)\,,~~~~\nonumber\\
&&
M_{V_2^+}^2\,\simeq\,\frac{1}{4}(g_V^2+g^2) F^2\,+\,\frac{1}{4}\frac{g_V^4}{g^2+g_V^2}f^2\,,
\eeqs
where $V_1^+ \simeq R^+$ and $V_2^+ \simeq (g W^+ - g_{V} L^+)/\sqrt{g^2 + g_{V}^2}$.

For the neutral gauge bosons, in addition to the massless photon and $Z$ boson,
 there are two heavy states $V_1^0$ and $V_2^0$ with masses nearly degenerate with their charged 
 counterparts in the parameter range of interest.
 We exhibit here only an approximate expression for the mass of $V_1^0$, valid for $f\ll F$:
 \beqs
 M_{V_1^0}^2&=&\frac{1}{4}\left[g_V^2(F^2+f^2)+g^{\prime\,2}(F^2-f^2)\right]\,,
 \eeqs
 where approximately $V_1^0 \simeq (g' B - g_{V} R^{0}) /\sqrt{g'^2 + g_{V}^2}$.

To describe fermion masses in our model, we note that the combination $\Phi_{L} \Phi \Phi_{R}$ transforms 
               as the Higgs field in the standard model. Hence we include the terms 

 \beqs
{\cal L}_{f}&=&-\,\frac{1}{\sqrt{2}F^2}\bar{q}_L\Phi_L\Phi \Phi_Rt_qq_R\,
-\,\frac{1}{\sqrt{2}F^2}\bar{\ell}_L\Phi_L\Phi\Phi_Rt_{\ell}\ell_R\,+\,{\rm h.c.}\,,
\eeqs
  where $
t_q\,=\,\left(\begin{array}{cc}
y_t & 0\cr
0 & y_b\end{array}\right)$,
$
\,t_{\ell}\,=\,\left(\begin{array}{cc}
y_{\tau} & 0\cr
0 & y_3\end{array}\right)\,
$ and we show explicitly only the third standard-model family. 
There are also  fermion kinetic terms with covariant derivatives as dictated by the quantum numbers of Table~\ref{Fig:4sites}. Having  imposed the nonlinear constraints on $\Phi_L$ and $\Phi_R$ and working in unitary gauge, these  terms           yield directly fermion mass expressions such as $m_t = y_{t} f/\sqrt{2}$ along with corresponding 
                formulae for the bottom quark, $\tau$ lepton and third-generation neutrino $\nu_3$. 
                The relation between the VEV $f$ and the electroweak scale will be 
                described in the next section. The generalization to three standard-model families with CKM flavor mixing 
                is straightforward. There remains one scalar field $h$ in the spectrum with mass $m_{h}^2 = 2 \lambda f^2$, 
                which we identify as the the particle discovered  by ATLAS~\cite{ATLASH} and CMS~\cite{CMSH}
                with mass $m_h \approx 125$ GeV.


The fine-tuning necessary to stabilize the Higgs mass in the standard model remains an issue also for
the current model. After determining the allowed range of parameters in the EFT,
we conclude that the amount of fine-tuning needed is no worse than in the standard model itself.

\section{Bounds  from Standard Model  Processes}
\label{Sec:SM}

We first discuss constraints from electroweak precision measurements.
We then turn to constraints arising from the
coupling of the Higgs boson to $W$ pairs and from its decay to two photons.

Electroweak precision parameters can be discussed conveniently by
examining the  low-energy EFT 
written in terms of
new  gauge fields $\bar{V}^i = (\bar{W}^{1}, \bar{W}^{2}, \bar{W}^{3}, \bar{B})$
and their propagators, obtained by expanding about $q^2 = 0$
the two-point functions derived from Eq.~(\ref{Eq:Lag}). 
We focus on  the transverse
polarizations of the gauge bosons and on the coupling to the currents made of standard-model fermions:
\beqs
\label{Eq:SMEFT}
{\cal L}&=&\frac{P_{\mu\nu}}{2}\bar{V}^{i,\mu}(-q)\bar{\pi}_{ij}(q^2)\bar{V}^{j,\nu}(q)\,+\,\\
\nonumber
&&\frac{\bar{g}}{2}\left[\frac{}{}\bar{W}^{i\,\mu}(q)J_{L\,i\,\mu}(-q)
+\bar{W}^{i\,\mu}(-q)J_{L\,i\,\mu}(q)\right]
\,+\,\frac{\bar{g}^{\prime}}{2}\left[\frac{}{}\bar{B}^{\mu}(q)J_{Y\,\mu}(-q)+\bar{B}^{\mu}(-q)J_{Y\,\mu}(q)\right]\,,
\eeqs
where $P^{\mu\nu}\equiv g^{\mu\nu}-q^{\mu}q^{\nu}/q^2$.
The $\bar{\pi}_{ij}(q^{2})$ functions can be expressed 
in terms of the parameters  $g$, $g^{\prime}$, $g_{V}$, $f$ and $F$. 
For our purposes, we will retain their $q^2$-dependence only up to ${\cal O}(q^2)$.
We follow the conventions of Ref.~\cite{Barbieri}, except that we rescale the gauge fields 
(and the gauge couplings $\bar{g}$ and $\bar{g}^{\prime}$) such that 
$\bar{\pi}^{\prime}_{33}(0)\,=\,1\,=\,\bar{\pi}^{\prime}_{BB}(0)$. 

 All the information we need for universal precision electroweak constraints is  contained in the
 functions $\bar{\pi}_{ij}(q^2)$~\cite{Barbieri}: our model falls into this universal class because of the charge assignments
 of all the fields, in particular the fact that there are no direct couplings of the SM fermions to the 
 $SU(2)_{L^{\prime}}\times SU(2)_{R^{\prime}}$ gauge bosons.~\footnote{Generalizations 
 of the model in which fermions 
 couple directly to the new gauge bosons  require a more general formalism 
 for electroweak precision physics~\cite{Witeck}.}
The $\hat{S}$ parameter, related to the $S$ parameter by Peskin and Takeuchi~\cite{PT} and the $\alpha_1$ parameter of the 
EW chiral Lagrangian~\cite{EWCL} as
$
\hat{S}\,=\,-\bar{g}^2\alpha_1\,=\,\frac{\alpha}{4\sin^2\theta_W}S\,,
$ is defined with these conventions as 
\beqs
\hat{S}&=&\frac{\bar{g}}{\bar{g}^{\prime}}\,\bar{\pi}^{\prime}_{3B}(0)\,,
\eeqs
where $\bar{\pi}^{\prime}$ indicates derivative of $\bar{\pi}$ in respect to $q^2$.

 In the neutral sector,  the function $\bar{\pi}(q^2)$ can be extracted from the matrix-valued functions
$\pi^0(q^2)\,\equiv\,q^2\,\mathbb{I}_4\,-\,{\cal M}^2_0$, written in the basis $(W,L,R,B)$,
by  inverting $\pi^0(q^2)$, 
by retaining only the four corners of the result, by inverting again and finally by expanding in small-$q^2$.
In the charged sector $\pi^+(q^2)$  is obtained by restricting to the first 3 rows and columns of $\pi^0(q^2)$,
and the analogous $\bar{\pi}^+(q^2)$ for the charged sector is obtained by inverting the $11$ element of $1/\pi^+$,
and then again Taylor expanding in small-$q^2$ and truncating at ${\cal O}(q^2)$.

We will impose the bound $\hat{S}<0.0039$.~\footnote{This is the $3\sigma$
bound for a light Higgs from~\cite{Barbieri}, obtained in a global fit of all the precision parameters that includes 1-loop corrections
from loops of SM fields, $h$ and top quark in particular. The bound has to be taken as indicative since the loop-level analysis of SM radiative corrections, which must be included to establish an experimental limit on $\hat{S}$, will involve modified couplings.}
The other universal precision parameters are discussed elsewhere~\cite{Barbieri}, and we expect them to be suppressed 
 as long as $g_V\gsim1$. For
our purposes we need only to stress that the $T$ parameter does not receive important contributions, as
the new gauge bosons preserve the $SU(2)_R$ custodial symmetry that is only broken by $g^{\prime}$ in the gauge sector
(models in which custodial symmetry is not implemented can also be considered, as for instance suggested in~\cite{CDQS}).

The result for the $\hat{S}$ parameter is
\beqs
\hat{S}&=&
\frac{2 f^2 g^2 \left(f^2+F^2\right)}{{g_V}^2 \left(2 f^2+F^2\right)^2+g^2 \left(2
   f^4+2 f^2 F^2+F^4\right)}\,.
\label{Eq:Shat}
\eeqs
Generically, this is ${\cal O}(1)$, but we have already noted that to accommodate the requirement 
         $M_{V_{1}^{+}} \gsim 2000$ GeV, we must take either  $g^2 \ll g_V^2$ or $f^2 \ll F^2$. For $g^2\ll g_{V}^2$ this becomes 
\beqs
\hat{S}&\simeq&\frac{2f^2(f^2+F^2)g^2}{(2f^2+F^2)^2g_{V}^2}\,,
\label{Eq:Sstrong}
\eeqs
which can also be derived in an alternative way
that we summarize in Appendix~\ref{Sec:Weinberg},
while for $f^2\ll F^2$ it becomes
\beqs
\hat{S}&\simeq&\frac{2g^2f^2}{(g_V^2+g^2)F^2}\,.
\eeqs

In either limit, the expressions for the $\hat{S}$ parameter can be combined with 
earlier expressions for the masses of the vector bosons, to conclude that throughout the parameter space
$\hat{S}\lsim 2M_{W}^2/M_{V_{1,2}}^2$. Thus for $M_{V_{1,2}}\gsim 2$ TeV one finds $\hat{S}\lsim 0.003$.
This has to do with the way in which we built the model itself: due to our set-up, {\it all} effects of new physics 
are suppressed by ${\cal O}(M_W^2/M_{V_i}^2)$ coefficients.~\footnote{For the precision parameters $W$ and $Y$ defined in Ref.~\cite{Barbieri}, for $g_V\gsim 1$ we find the approximate relations $W\sim g^2/(4g_V^2)\hat{S}$ and $Y\sim g^{\prime\,2}/(4g_V^2)\hat{S}$, which make them further suppressed 
with respect to $\hat{S}$.}
The same applies to other coefficients of the electroweak chiral Lagrangian at ${\cal O}(p^4)$, for which the 
bounds are weaker~\cite{Fabbrichesi}.

From the expression for the charged $\bar{\pi}^+(q^2)$ we can extract the Fermi constant,
$G_F/\sqrt{2}\equiv1/(2v_W^2)$, where  $v_W\simeq 246$ GeV is given by
\beqs
v_W^2&=&\lim_{q\rightarrow 0}\frac{4}{\bar{g}^2}\left(q^2-\bar{\pi}^+\right)\,=\,\frac{f^2F^2}{F^2+2f^2}\,.
\label{Eq:VW}
\eeqs

The low energy effective theory includes also the Higgs $h$.
The coupling of the Higgs to the fermions $\psi$ is controlled by the fermion mass $m_\psi$
\beqs
{\cal L}_f&=&\cdots -c\frac{h}{v_W}m_{\psi}\bar{\psi}\psi\,,
\eeqs
related to  the SM prediction by the factor  
\beqs
c&\equiv& v_W/f\,.
\eeqs

The coupling of $h$ to two $W$ bosons 
is rescaled with respect to the standard-model coupling $M_W^2/v_W$ by a multiplicative factor $a$.
We find the approximate relations
\beqs
a&\simeq&\left\{\begin{array}{cc}
\frac{F^3}{(F^2+2f^2)^{3/2}}\,=\,c^3\,,& (g\ll g_V)  \vspace{2mm} \\ 
1\,-\,\frac{(2g^4+4g^2g_V^2+3g_V^4)f^2}{(g^2+g_V^2)^2 F^2}\,,& (f\ll F)
\end{array}
\right.\,,
\eeqs
and  analogous expressions hold for the coupling to two $Z$ bosons.
For any range of $g/g_V$, there is significant suppression of the 
Higgs coupling to the $W$ bosons, relative to the
        standard model, unless $f \ll F$.  We impose this restriction, 
ensuring compatibility with current data from ATLAS and
      CMS~\cite{ATLASCMS}.  Then, as we will see in Section~\ref{Sec:phenomenology}, maintaining a 
fixed ratio  $M_W / M_{V_{1}}$ leads to an
        upper bound on $g_V$.

A similar calculation yields the couplings $a_1$ and $a_2$
of the Higgs to the heavy vectors $V_1^\pm$ and $V_2^\pm$, all of which are suppressed. We
do not report them here, but we find that
these coefficients satisfy the sum rule $a+a_1+a_2=c$.
This allows us to write the decay rate of the Higgs to two photons approximately as
\beqs
\Gamma(h\rightarrow\gamma\gamma)&=&\frac{G_F\alpha^2m_h^3}{128\sqrt{2}\,\pi^3}
\left|c\,A_t(\tau_t)N_cN_fQ^2\,+\,a\,A_W(\tau_W)\,+\,(c-a)A_W(0)\right|^2\,,
\label{Eq:gammagamma}
\eeqs
where $A_t(\tau_t)\simeq 1.38$, $A_W(\tau_W)\simeq -8.3$, and $A_W(0)=-7$. 
The three terms represent the contribution of the top loop (which implies $N_c=3$, $N_f=1$ and $Q^2=4/9$),
of the $W$ loops and the loops of heavy charged vectors, respectively.
For $a=c=1$ this is the SM rate.

\section{Decay Processes of Heavy Vectors}
\label{Sec:decay}

We take the masses of the heavy vector bosons of our EFT to be at least as large as suggested by the ATLAS and CMS di-boson
excesses~\cite{ATLAS,CMS}. Depending on parameter values, the masses of the six particles can be well split or nearly degenerate.
In either case the dominant decays will be two-body, to either a pair of SM fermions or a pair of SM bosons. 
To determine the fermionic decay width of the charged vector states, we start from the 
full $\pi^+(q^2)$ matrix function in the $(W,L,R)$ basis of our EFT, 
and make use of the replacement
\beqs
&&\left(\frac{1}{q^2-M_W^2}\right)_{\rm SM}\,\rightarrow\,
\left(\frac{1}{\pi^+(q^2)}\right)_{WW}\,\label{Eq:2points}\\ &&\equiv\,\nonumber
\frac{4 \left(F^2 g_{V}^2-4 q^2\right) \left[g_{V}^2 \left(2
   f^2+F^2\right)-4 q^2\right]}{f^2 g_{V}^2 \left[-g^2g_{V}^2F^4+8 F^2 q^2 \left(g^2+g_{V}^2\right)-32 q^4\right]+4
   q^2 \left(F^2 g_{V}^2-4 q^2\right) \left[F^2
   \left(g^2+g_{V}^2\right)-4 q^2\right]}\,,\\
   &&\simeq\frac{r_W}{q^2-M_W^2}\,+\,\frac{r_{V_1^+}}{q^2-M_{V_1^+}^2}\,+\,\frac{r_{V_2^+}}{q^2-M_{V_2^+}^2}\,,
\eeqs
in the relevant amplitude.
The residues are approximately given by the following expressions, valid  for $g\ll g_V$:
\beqs
r_W&\simeq&1\,-\,\frac{g^2}{g_{V}^2}\left[1-\frac{2f^2(f^2+F^2)}{(2f^2+F^2)^2}\right]\,,~~
r_{V_1^+}\,\simeq\,\frac{g^2}{2g_{V}^2}\,,~~
r_{V_2^+}\,\simeq\,\frac{g^2F^4}{2g_{V}^2(2f^2+F^2)^2}\,,
\eeqs
while in the limit $f\ll F$ we have
\beqs
r_W&\simeq&\frac{g_V^2}{g_V^2+g^2}+\frac{2f^2g^2g_V^4}{F^2(g^2+g_V^2)^3}\,,
~~
r_{V_1^+}\,\simeq\,\frac{g_V^2f^4}{g^2F^4}\,,~~
r_{V_2^+}\,\simeq\,\frac{g^2}{g_V^2+g^2}-\frac{2f^2g^2g_V^4}{F^2(g^2+g_V^2)^3}
\,.\label{Eq:residuesf}
\eeqs
Both satisfy the sum rule $1=r_W+r_{V_1^+}+r_{V_2^+}$.
These approximations have to be used carefully:  the two limits do not commute,
as for instance taking the $g\rightarrow 0$ limit of the second set of approximations would yield an incorrect result.

The partial width of the charged gauge bosons into standard-model fermions can be obtained directly from the 
corresponding width of the $W$ boson, with three modifications: the mass of the gauge bosons has to be replaced, 
the coupling is suppressed by $\sqrt{r_V}$,
and all heavy fermions have to be included. 
The result for the decay of the charged heavy vectors to $e^+\nu_e$  is
\beqs
\Gamma\left(V_{1,2}^+\rightarrow e^+\nu_e\right)&=&r_{V_{1,2}^+}\,\frac{g^2M_{V_{1,2}^+}}{48\pi}\,,
\eeqs
and  the total decay rate to SM fermions is
$
\Gamma\left(V_{1,2}^+\rightarrow \psi^\prime\bar{\psi}\right)\,=\,3(1+N_c)\Gamma\left(V_{1,2}^+\rightarrow e^+\nu_e\right)
$.
In the case of the neutral vectors, the analogue of Eq.~(\ref{Eq:2points}) involves mixing matrices.
We do not report the details, but we perform an exact numerical study later in the paper.

 Because the masses of all the new spin-one states 
are much larger than the masses of the electroweak gauge bosons, the Goldstone boson equivalence theorem applies.
Decay rates involving $W$ and $Z$ 
are dominated by the contribution of the longitudinally polarized particles,
while we neglect the contribution of transverse polarizations.
In a somewhat similar manner, the fact that the Higgs particle has a mass at least $15$ times lighter than the new spin-one states 
means that decay rates involving the Higgs particle can be computed (at leading order) by setting $f=0$.
For these reasons we estimate the 
 di-boson decay rates with $f=0$, so that the relatively small amount of electroweak symmetry breaking is neglected.

The electroweak bosons in this limit are massless, 
and we  reinstate all the degrees of freedom of $\Phi$. We write
\beqs
\Phi&=&\left(\begin{array}{cc}
h_0+ih_3 & ih_1+h_2\cr
 ih_1-h_2 & h_0 -i h_3 \end{array}\right)\,,
\eeqs
with canonically normalized, real $h_i$.
The couplings of the neutral,  heavy spin-one states to the scalars are
 \beqs
 {\cal L}_{b}&=&\cdots\,+\frac{g_{V}}{\sqrt{2}}
 \left[ \frac{R^3+L^3}{\sqrt{2}}(h_1\partial h_2-h_2\partial h_1)
 +\frac{R^3-L^3}{\sqrt{2}}(h_0\partial h_3-h_3\partial h_0)+\cdots\right]\,.
 \eeqs
 In these approximations,   the partial width
 (at leading order in $f\ll F$) is given by
 \beqs
 \Gamma(V_{1}^+\rightarrow \mbox{ di-bosons})&\simeq&\frac{g_V^2}{96\pi}M_{V_{1}^+}\,.
 \eeqs
 For the other vector bosons, even in the $f=0$ limit we must retain the mixing due to the gauge couplings, and find
 accordingly that
 \beqs
 \Gamma(V_{2}^{+\,,\,0}\rightarrow \mbox{ di-bosons})&\simeq&\left(\frac{g_V^2}{g_V^2+g^2}\right)\frac{g_V^2}{96\pi}M_{V_{2}^{+,0}}\,,
 \eeqs
 for both the charged and neutral vectors, and for the neutral state $V_1^0$
  \beqs
 \Gamma(V_{1}^0\rightarrow \mbox{ di-bosons})&\simeq&\left(\frac{g_V^2}{g_V^2+g^{\prime\,2}}\right)\frac{g_V^2}{96\pi}M_{V_{1}^0}\,.
 \eeqs

When the mass splitting is large enough, there exist also decays of the type $V_2\rightarrow V_1+h,X$, 
where $X$ is a SM gauge boson. These decays contribute negligibly to the total width.

\section{Fine Tuning}
\label{Sec:finetuning}
Before presenting phenomenological and numerical estimates, we discuss briefly the amount
of fine tuning intrinsic to our model. As with any EFT, this discussion can provide only general
guidance on the issue of fine tuning and the scale of new physics. A specific UV completion
could modify the discussion. Nevertheless, an EFT-based discussion has the virtue
of generality and it reveals some interesting features.
The tree-level 
scalar potential reproduces the standard model.
The general form of the one-loop potential in the external field language is
\beqs
V_1&=&\frac{\Lambda^2}{32\pi^2}{\cal ST}r\,{\cal M}^2\,+\,\frac{1}{64\pi^2}
{\cal ST}r\,\left[({\cal M}^2)^2\left( \ln\frac{\,{\cal M}^2}{\Lambda^2}\,+\,c_i\right)\right]\,,
\eeqs
where $\Lambda$ is the UV cutoff, ${\cal ST}r$ is a trace in which fermionic degrees of freedom have negative weight,
${\cal M}^2(h)$ is the second derivative of the interaction part of the Hamiltonian, evaluated for general
$h$, and $c_i$ are scheme-dependent and field-dependent constants.

In the standard model, 
the dominant quadratically divergent part of the potential contributes to the 
Higgs mass~\cite{Einhorn:1992um}
\beqs
\Delta m_h^2({\rm SM})&=&\frac{\Lambda^2}{32\pi^2}
\left[\frac{3}{2}(3g^2+g^{\prime\,2})+\frac{6m_h^2}{v_W^2}-\frac{24m_t^2}{v_W^2}\right]\,+\cdots\,.
\eeqs
For $m_h\simeq 125$ GeV, and $\Lambda\simeq 10-30$ TeV, this results in
$|\Delta m_h^2| \simeq 2.5-25$ TeV$^2$. This necessitates 
a counter-term chosen with  
fine-tuning ${\cal Z}\equiv | m_h^2/\Delta m_h^2|\simeq 0.006-0.0006$. 

In our model, for $f\simeq v_W$:
 \beqs
\Delta m_h^2&\simeq&\frac{\Lambda^2}{32\pi^2}\left(9g_V^2+\frac{6m_h^2}{v_W^2}
-\frac{24m_t^2}{v_W^2}\right)\,+\,\cdots\,.\label{Eq:finetuning}
\eeqs
Notice that $g$ and $g^{\prime}$ do not appear at this order, while they do 
in the log-divergent and finite corrections.

A comparison of these two expressions reveals that for $g_V \simeq g $, the magnitude and sign of
required fine tuning are the same. For larger $g_V$, the possibility of some accidental cancellation
with the top-quark term arises in our EFT. It would be almost exact for $g_V \simeq 1.05$. For the
range $g_V \geq 2.0$, to emerge from our phenomenological study in the next section, the first
term in Eq.~(\ref{Eq:finetuning}) dominates, reversing the sign of the quadratically divergent contribution
to $\Delta m_h^2$. For $g_V$ not too far above $2.0$, the magnitude of the required fine
tuning is approximately the same as in the standard model.

\section{Phenomenology and Numerical Study}  
\label{Sec:phenomenology}

\begin{figure}[h]
\begin{center}
\begin{picture}(400,130)
\put(-10,5){\includegraphics[height=4.5cm]{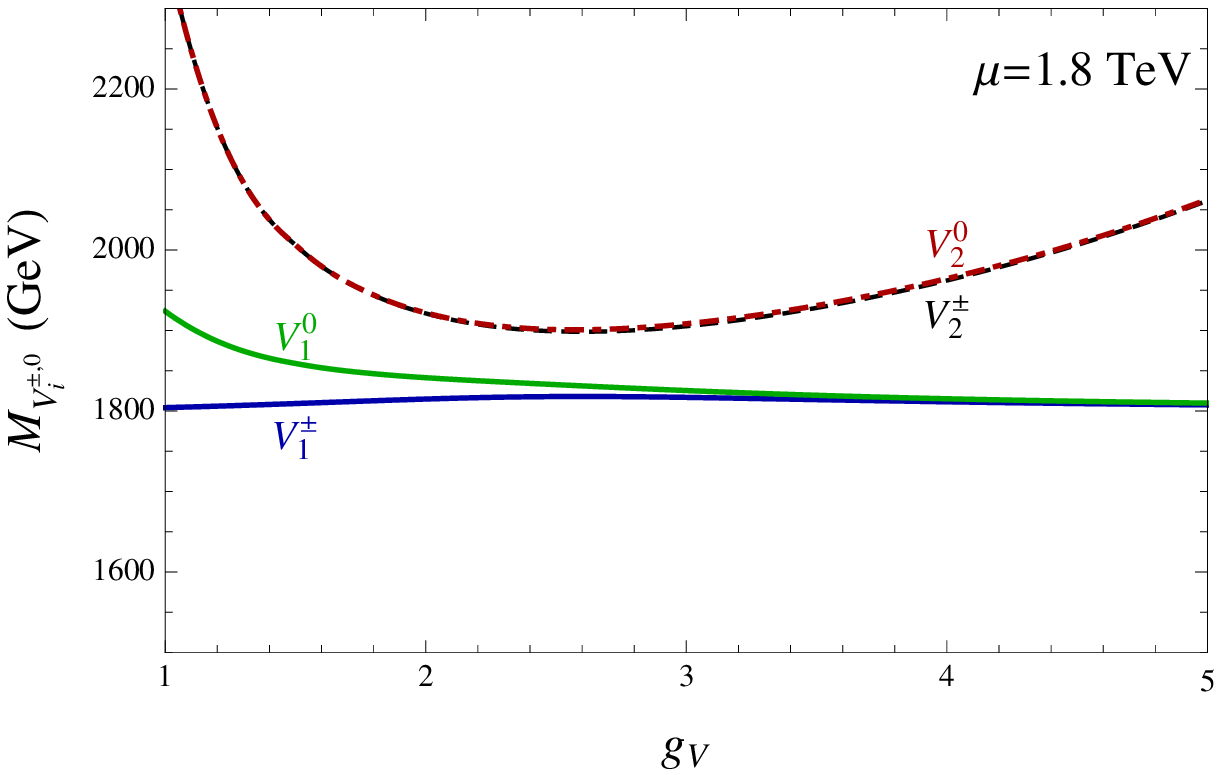}} 
\put(210,5){\includegraphics[height=4.5cm]{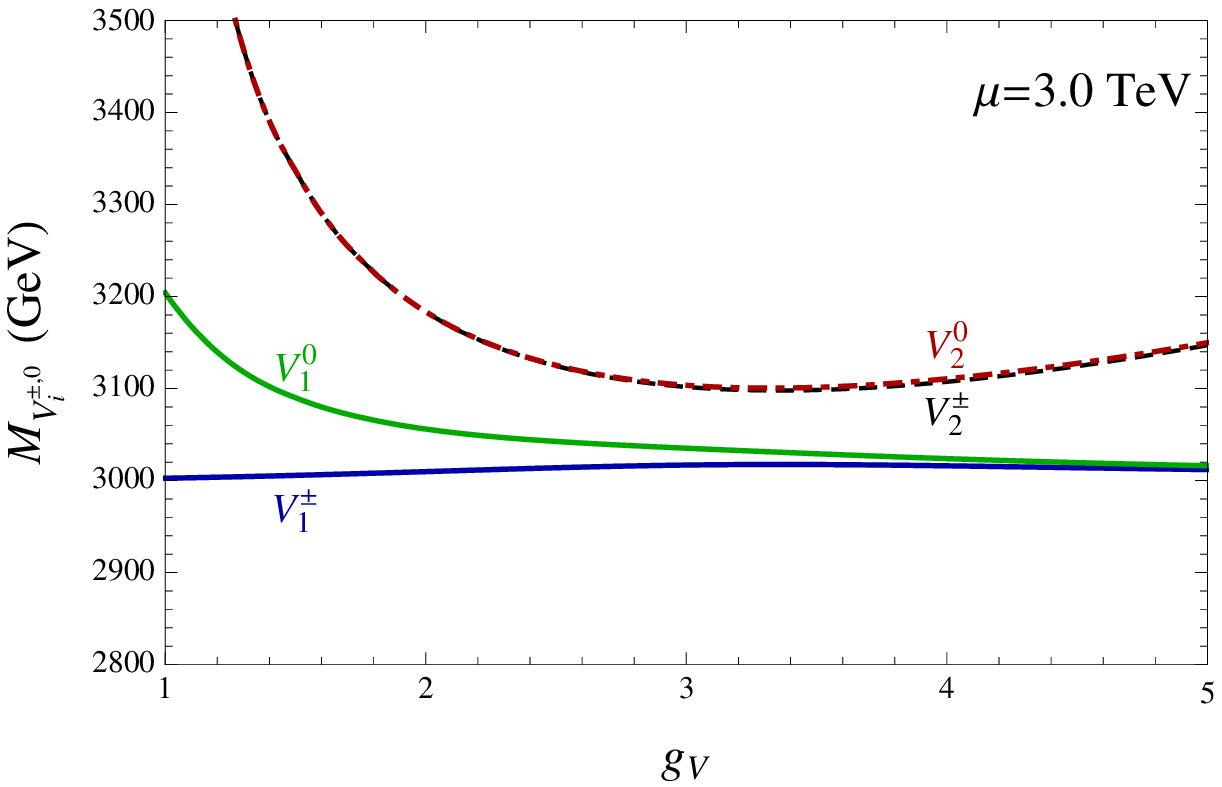}}
\end{picture} 
\caption{Mass spectrum (in GeV) of the heavy vector bosons in the model, as a function of $g_V$, for the choices of parameters discussed in the text.
The left panel shows the masses for $\mu=1.8$ TeV, the right panel for $\mu=3.0$ TeV.
The blue continuous curves show the mass of $V_1^{\pm}$, the black dashes line the mass of $V_2^{\pm}$, the
green continuous lines show the mass of $V_1^0$ and the red dot-dashed line the mass of $V_2^0$.}
\label{Fig:spectrum}
\end{center}
\end{figure}

We first look at the constraints on the model from precision electroweak physics and from 
the properties of the Higgs particle, and then at the signatures of the new particles in direct searches
at high energies.
We are mostly interested in the regime $f\ll F$, 
 in which limits from electroweak precision tests are satisfied,
 in which the Higgs couplings are close to the standard model ones, 
and in which fine-tuning is at a comparable level with the standard model. 

There are five parameters, $g$, $g^{\prime}$, $g_V$, $f$ and $F$. We
 keep fixed the indicative scale of the new particles $\mu\equiv\frac{1}{2}g_VF$.
For each choice of $g_V$ we then use Eq.~(\ref{Eq:VW}), together with the exact diagonalization of the
mass matrices in order to fix the parameters $f$, $g$ and $g^{\prime}$
so that we reproduce the standard model values $M_W\simeq 80.4$ GeV, $M_Z\simeq 91.2$ GeV, $v_W\simeq 246$ GeV.
With all of this in place, all the new physics depends on only one parameter, namely $g_V$.
In the following, we make
 two representative choices  $\mu=1.8$ TeV and $\mu=3.0$ TeV.
 
We compute the spectrum  by diagonalizing numerically the mass matrices.
The results are shown in Fig.~\ref{Fig:spectrum}. Notice that $V_2^\pm$ and $V_2^0$ are so close in mass that
they appear as just one line (the heaviest mass).
Their masses come close to the masses of the lighter $V_1$ bosons for intermediate values of $g_V$,
but differ substantially both for large $g_V$ and small $g_V$. 
While for large $g_V$ one can choose $f$ and $F$
to be of similar order, when $g_V$ becomes smaller one is forced towards the $f\ll F$ limit
 in which the masses of $V_1$ and $V_2$  become degenerate.
 But having kept $\mu$ fixed, as well as having imposed the requirement that the light masses agree 
 with the standard-model bosons,
  the limit of small $g_V$ is a limit in which the explicit breaking of the left-right 
 symmetry due to the $g$ and $g^{\prime}$ couplings is enhanced.

The mass of $V_1^\pm$ is approximately degenerate with $V_1^0$ for large $g_V$, but
the degeneracy is lifted at small $g_V$.
This charged-neutral splitting is due to the fact that the diagonalization of the mass matrix for the neutral 
vectors differs by effects that are controlled by the $g^{\prime}$ coupling. These effects are hence negligible when $g_V\gg g^{\prime}$,
but are enhanced when $g_V$ is small.

\subsection{Constraints from Indirect Searches}
\label{Sec:indirect}

\begin{figure}[h]
\begin{center}
\begin{picture}(400,260)
\put(-10,135){\includegraphics[height=4.5cm]{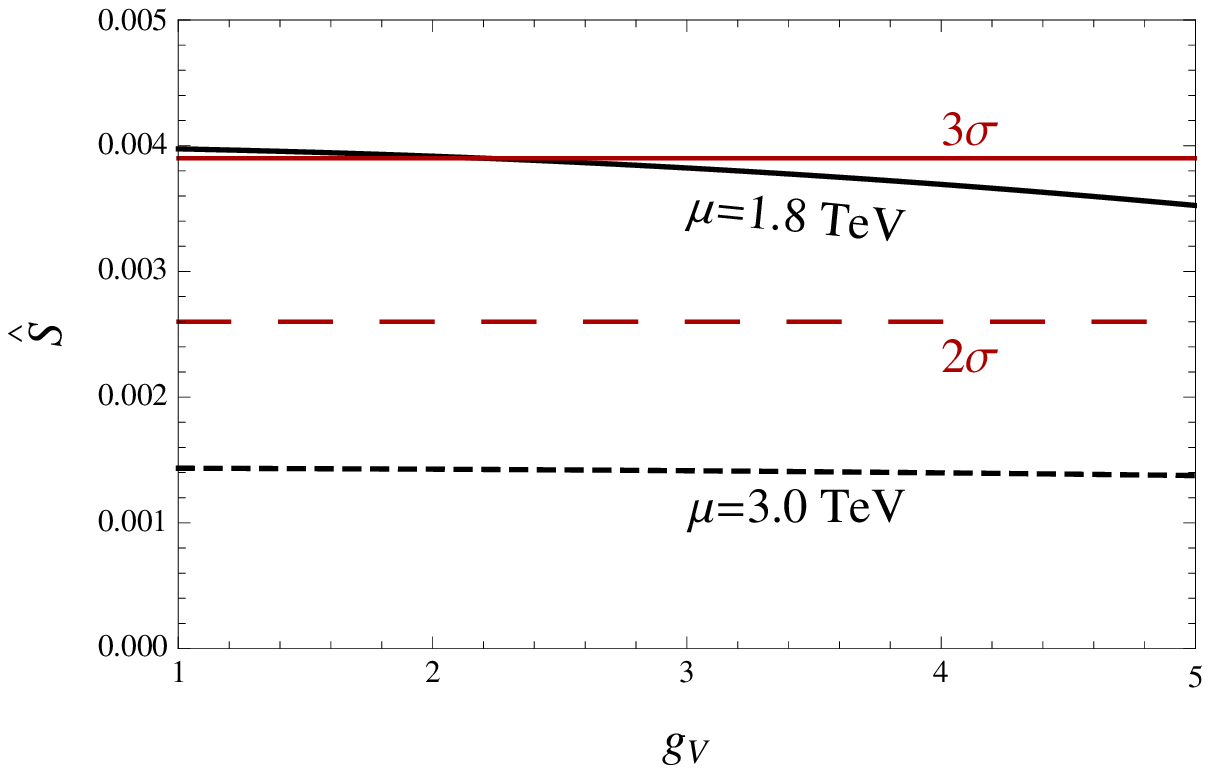}}
\put(210,135){\includegraphics[height=4.5cm]{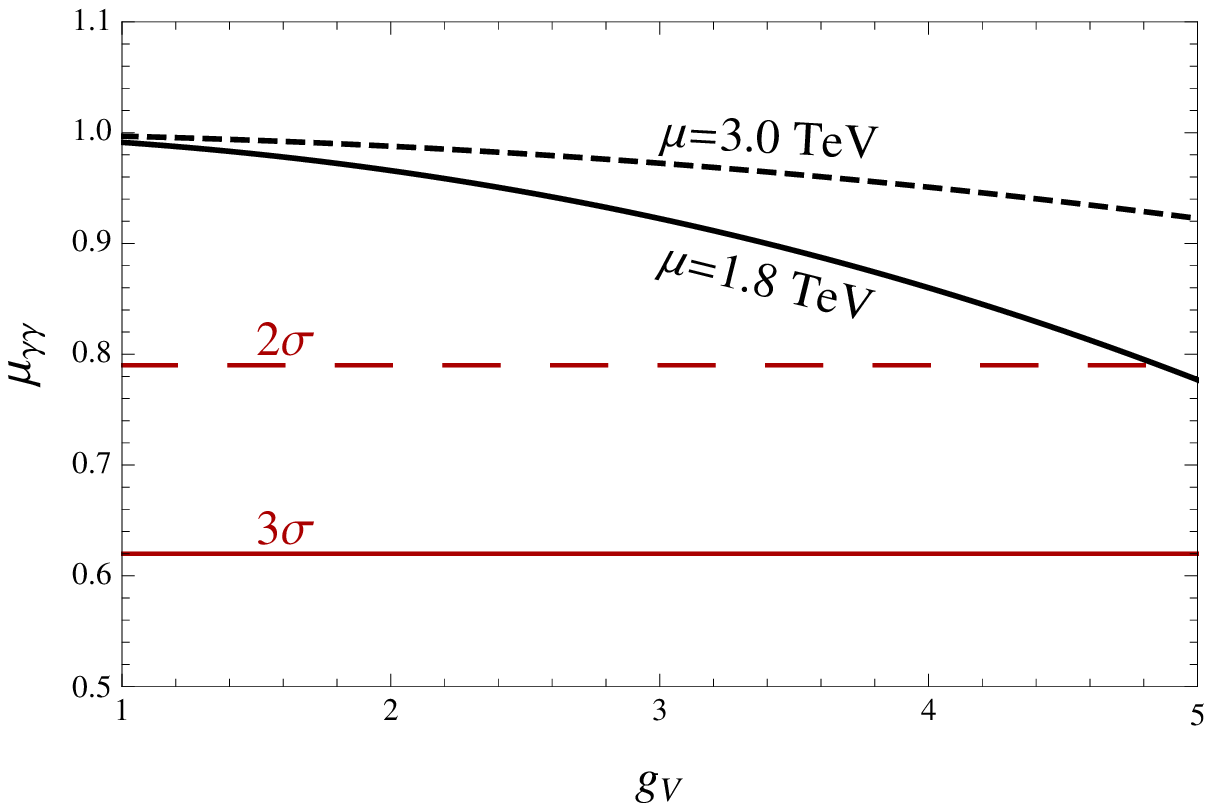}}
\put(110,5){\includegraphics[height=4.5cm]{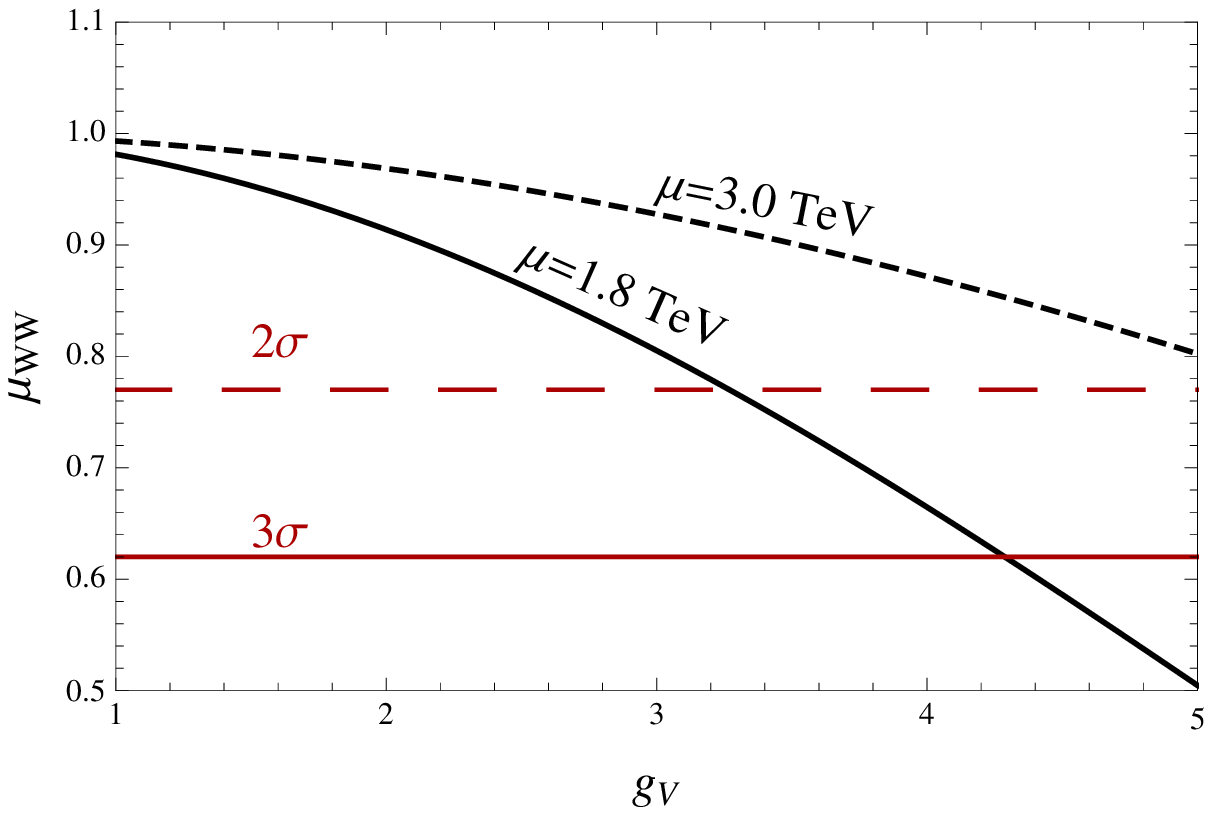}}
\end{picture} 
\caption{Bounds from indirect searches. The top left panel show $\hat{S}$ as a function of $g_V$, and for the choices of parameters described in the text (in black). 
The top right panel shows $\mu_{\gamma\gamma}$ for the same choices of parameters
and the last panel shows $\mu_{WW}$.
The continuous lines are for $\mu=1.8$ TeV, the dashed lines for $\mu=3.0$ TeV.
In all plots we compare to the $3\sigma$ (red) and $2\sigma$ (red long dashing)
  indicative bounds discussed in the text.}
\label{Fig:indirect}
\end{center}
\end{figure}

  For the electroweak precision parameter $\hat{S}$, we use Eq.~(\ref{Eq:Shat}), and we show the results in the top left panel of Fig.~\ref{Fig:indirect}.
We also look at the couplings of the Higgs $h$, making use of the exact result for $c=v_W/f$
and  numerical results for $a$.

The measurements of cross sections times decay rates of processes involving the Higgs place important bounds on our model parameter space. The processes in which a Higgs decays to $W$ bosons or photons are particularly important. To derive these bounds,
we look at the signal strength as measured by the ATLAS and CMS collaborations in Run 1.
The signal strength $\mu_i$ is defined as the number of observed events of a given type $i$, 
normalized relative to the prediction from the standard model computed with Higgs mass $125-126$ GeV.
Thus $\mu_i=1$ indicates perfect agreement with the standard model. The use of the signal strength originates from the fact that the combined $2$-parameter fits in~\cite{ATLASCMS}, cannot be used, as our expression for the $h\rightarrow \gamma\gamma$ rate contains contributions from the heavy vector bosons, which are absent in the combination done by the experimental collaborations.

We compare the signal strength $\mu_{\gamma\gamma}$ to the weighted average of the 
signal strengths measured by CMS and ATLAS, that we find to be
 $\mu_{\gamma\gamma}=1.13\pm 0.17$ from~\cite{ATLASCMS}, 
by making use of the rough approximation
\beqs
\mu_{\gamma\gamma}&\simeq&\frac{c^2}{0.75\,c^2\,+\,0.25\,a^2}\,\frac{\Gamma(h\rightarrow\gamma\gamma)}{\Gamma(h\rightarrow \gamma\gamma)_{\rm SM}}\,,
\eeqs
together with Eq.~(\ref{Eq:gammagamma}). 

This approximation for $h\rightarrow \gamma\gamma$ has the following origin (see also Appendix~\ref{Sec:Higgs}).
The factor of $c^2$ arises from the fact that the dominant production cross section comes from
gluon-gluon fusion ($ggF$), in which the coupling of the Higgs particle to the gluons is due to a loop of top quarks. The coupling of the Higgs to top quarks is suppressed in our model by a factor of $c$ with respect to the standard model. The denominator is an approximation of the rescaling of the total width of the Higgs particle in our model: 
neglecting the small contribution from $\gamma\gamma$ 
itself. For $m_h\sim 125$ GeV in the standard model the branching ratio (BR) is approximately $75\%$ in $bb$, $cc$, $\tau\tau$ and $gg$, all of which are 
suppressed as $c^2$, while $25\%$ comes from decays into $WW^{\ast}$  that are 
suppressed by $a^2$ with respect to the standard model, together with a smaller  $ZZ^{\ast}$ fraction, suppressed in a similar way. 
Finally the ratio of the $\gamma\gamma$ rate with respect to the standard model has been discussed earlier in the paper. 
Here and in the following $c$ and $a$ are computed numerically,
without any of the approximations we discussed in Section~\ref{Sec:SM}.

Finally, we compare the signal strength $\mu_{WW}$ to the weighted average of CMS and ATLAS
$\mu_{WW}=1.07\pm 0.15$ from~\cite{ATLASCMS},  by making use of the rough
approximation
\beqs
\mu_{WW}&\simeq&\frac{c^2a^2}{0.75\,c^2\,+\,0.25\,a^2}\,.
\eeqs

The results of both analysis are shown in Fig.~\ref{Fig:indirect},  from which we deduce
the bound $g_V\lsim 4.2$ at $3\sigma$.
The behavior of the curves follows from the fact that for $f/F\rightarrow 0$ our EFT 
coincides with the standard model. In this limit $f\rightarrow v_W$ and hence $c=1$. 
Because we keep $\mu=\frac{1}{2}g_VF$ fixed, this is the limit in which $g_V$ becomes small.
By contrast, taking $g_V$ large for fixed $\mu$ means lowering $F$ to the point where the modifications of the Higgs couplings
become large, conflicting with the experimental bounds.

\begin{figure}[h]
\begin{center}
\begin{picture}(400,140)
\put(-10,5){\includegraphics[height=4.5cm]{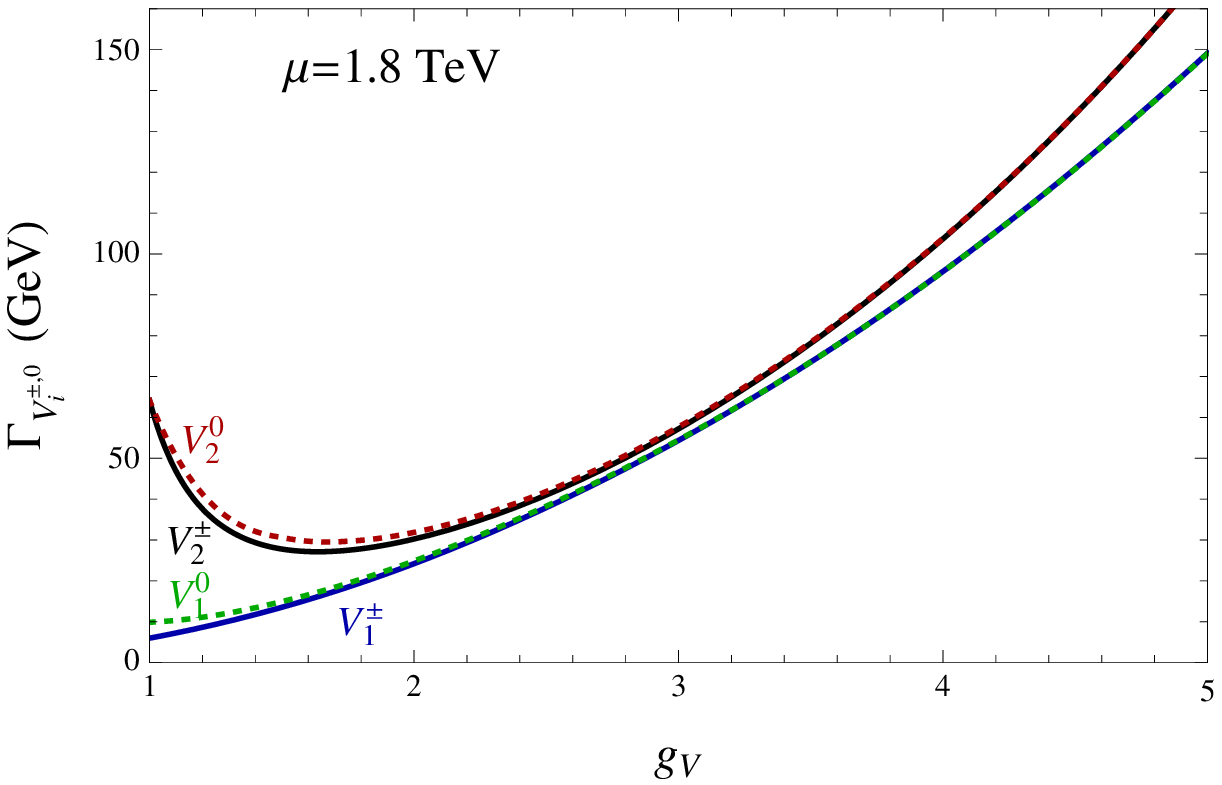}}
\put(210,5){\includegraphics[height=4.5cm]{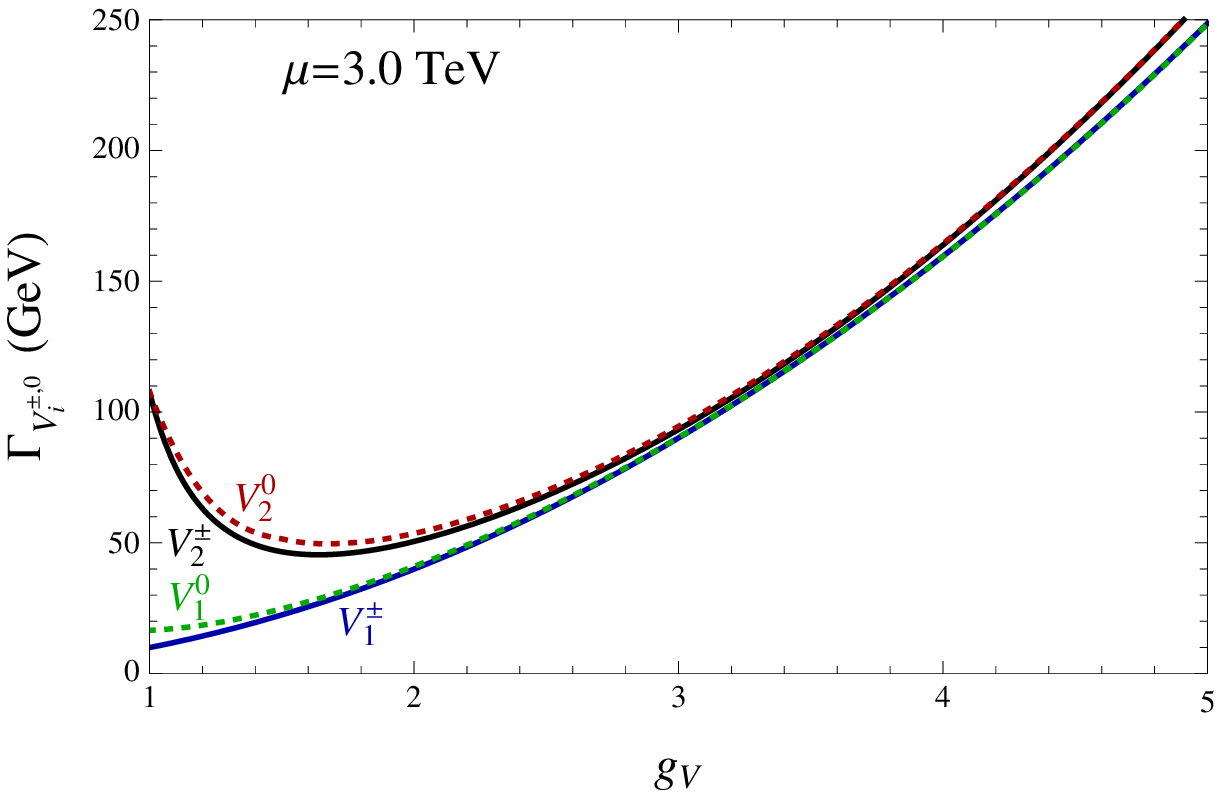}}
\end{picture} 
\caption{The total widths (in GeV) of the charged particles $V_1^{\pm}$ (blue), $V_2^{\pm}$ (black). $V_1^0$ (dashed, green)
and $V_2^0$ (dashed, red). 
The left panel shows $\mu=1.8$ TeV, the right panel $\mu=3.0$ TeV.}
\label{Fig:Gamma}
\end{center}
\end{figure}

\subsection{Direct Searches}
\label{Sec:direct}

We next study the strength of the coupling to the currents $J_L$ and $J_Y$.
We do so by numerically computing the residues at the poles in the propagators for charged and neutral gauge bosons,
as in Eq.~(\ref{Eq:2points}).
The results affect the production cross sections, as well as the partial widths
of the particles. We do not report these intermediate results, 
except for commenting on the fact that we checked explicitly that 
the couplings of the physical $W$ and $Z$ bosons reproduce the standard-model values, within 
the accuracy required by precision physics. 

The widths of the heavy particles are shown in
 Fig.~\ref{Fig:Gamma}. The total widths include all the di-boson channels (computed 
with the equivalence theorem in 
Section~\ref{Sec:decay}) as well as the decay to all standard-model fermion pairs. The total widths
range from tens of GeV to $100-200$ GeV. 
Because  the width of the neutral particles is close to that of their charged partners, we focus our attention on the charged particles
in this discussion.
The width of the $V_{1}^+$
decreases as $g_V$ is reduced for two reasons. Firstly, its bosonic width
is proportional to $g_V^2$. Secondly, in the small $g_V$ limit, the quantity
$f/F$ becomes small, and the  $V_{1}^+$ becomes predominantly the
$R^+$ state, as discussed in Section~\ref{Sec:EFT}. Its coupling to fermion pairs is
suppressed by the factor $r_{V_{1}^+}$ shown in Eq.~(\ref{Eq:residuesf}). 
The bosonic width of the $V_{2}^+$ also decreases with $g_V$ but then the width to
fermion pairs kicks in. The mass eigenstate $V_{2}^+$ becomes a linear combination of
$L^+$ and $W^+$ in this limit, and its fermionic width is not suppressed.
This can be seen from the factor $r_{V_{2}^+}$ in Eq.~(\ref{Eq:residuesf}). As a result, the
width of  $V_{2}^+$ has a minimum around $g_V \approx 1.5$.

\begin{figure}[h]
\begin{center}
\begin{picture}(440,290)
\put(0,0){\includegraphics[height=4.5cm]{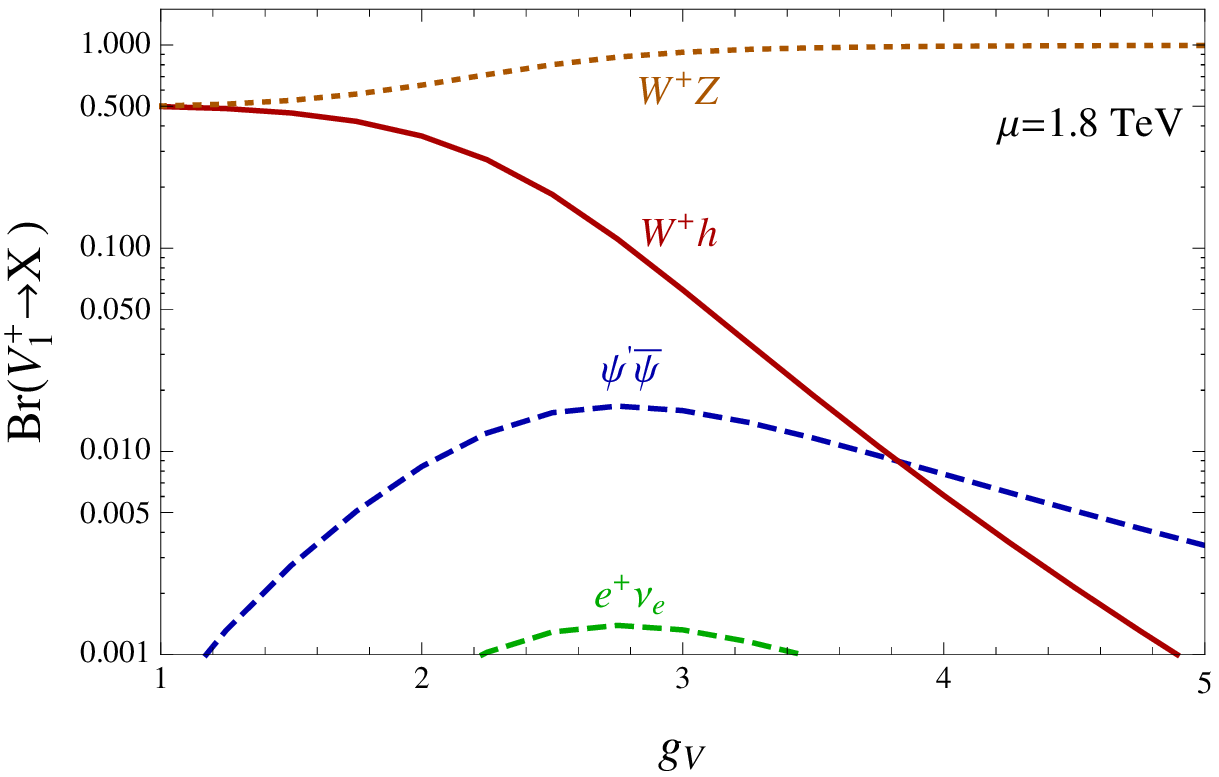}}
\put(220,0){\includegraphics[height=4.5cm]{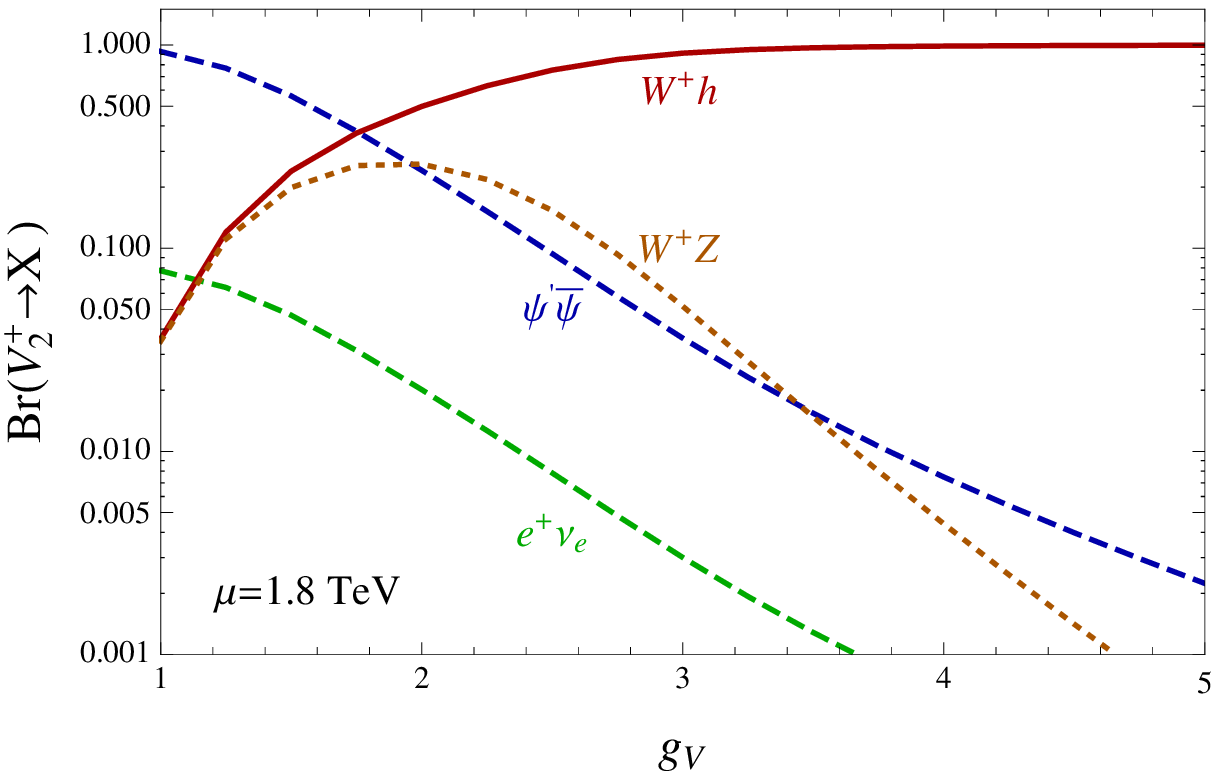}}
\put(0,145){\includegraphics[height=4.5cm]{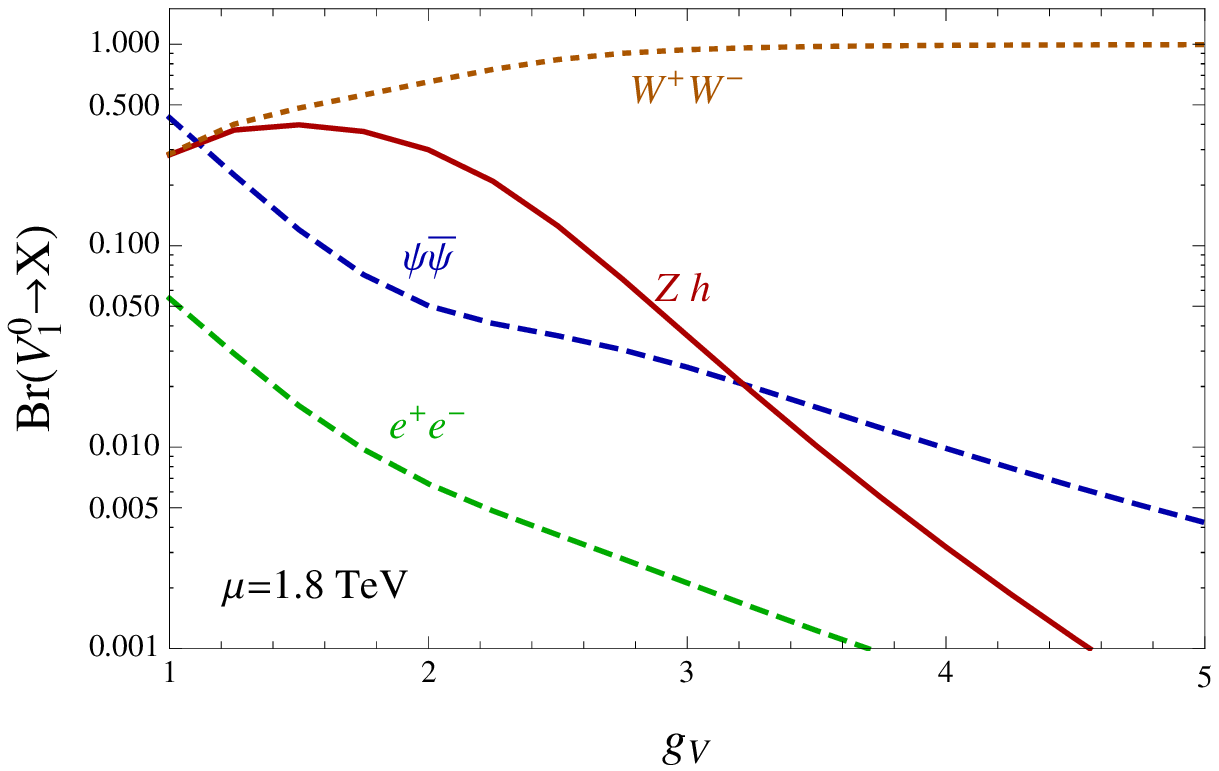}}
\put(220,145){\includegraphics[height=4.5cm]{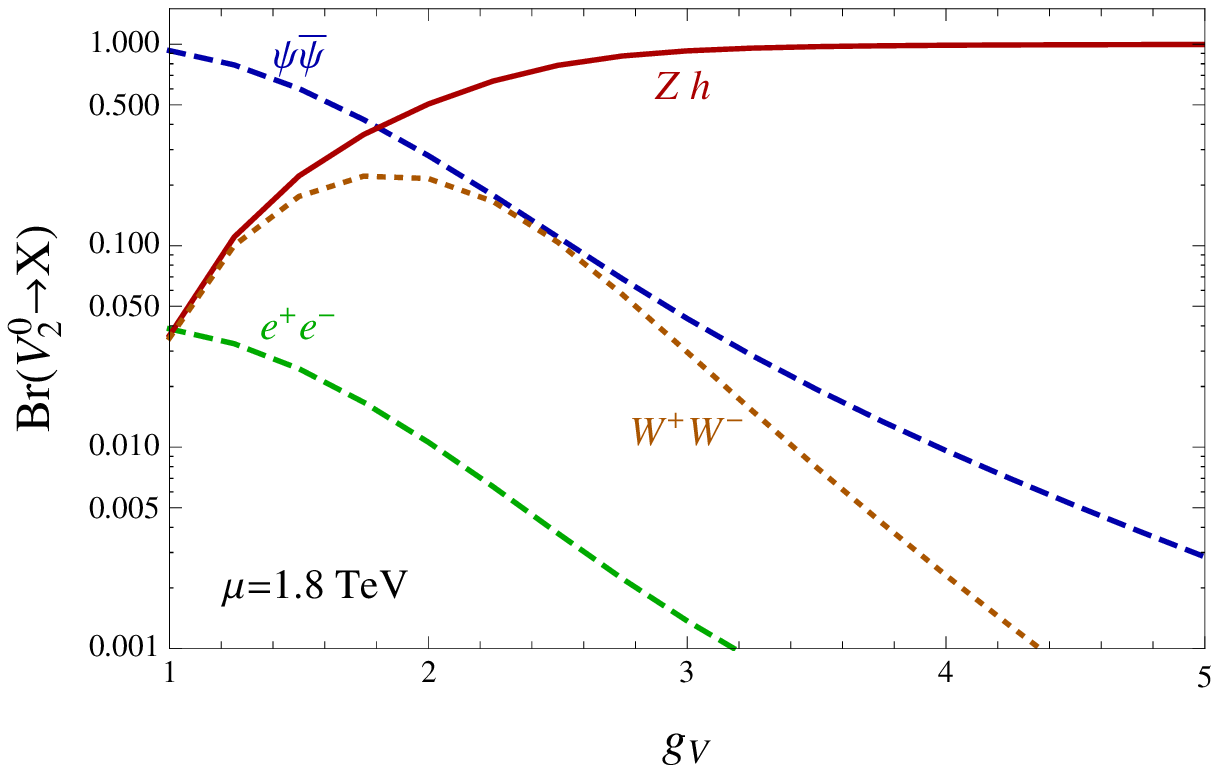}}
\end{picture} 
\caption{Branching fractions for $\mu=1.8$ TeV. Top to bottom, left to right,
the branching ratios of $V_1^0$, $V_2^0$, $V_1^+$ and $V_2^+$.}
\label{Fig:BR}
\end{center}
\end{figure}

To provide more detail, we next look at the branching ratios of all the heavy particles, including the neutral ones,  in Fig.~\ref{Fig:BR} for $\mu=1.8$ TeV.  The branching fractions for $\mu=3$ TeV have similar behavior and we do not show them. The suppressed branching fractions of the heavy neutral and charged $V_2$ particles to $V_1$ particles and SM bosons have not been included in Fig.~\ref{Fig:BR}. For the $V_{1}^+$,  the di-boson channels dominate the decay width throughout the exhibited $g_V$ range due to the suppressed coupling to standard model fermions discussed above. For the $V_{1}^0$, the standard-model fermion channel is generally suppressed but becomes comparable to the di-boson channel for very small $g_V$. The reason is that in this limit the $V_{1}^0$ becomes a linear combination of $R^0$ and $B$, both of which couple to SM-fermion pairs with electroweak strength. For both the $V_{2}^0$ and $V_{2}^+$, which become linear combinations of $L$ and $W$ in the small-$g_V$ limit, the SM-coupling to the fermions is somewhat stronger than for the $V_{1}^0$, so these modes dominate below $g_{V} \approx 1.5$. 
 Each of these features has important consequences for the various production cross sections. 

\begin{figure}[h]
\begin{center}
\begin{picture}(480,270)
\put(0,140){\includegraphics[height=4.5cm]{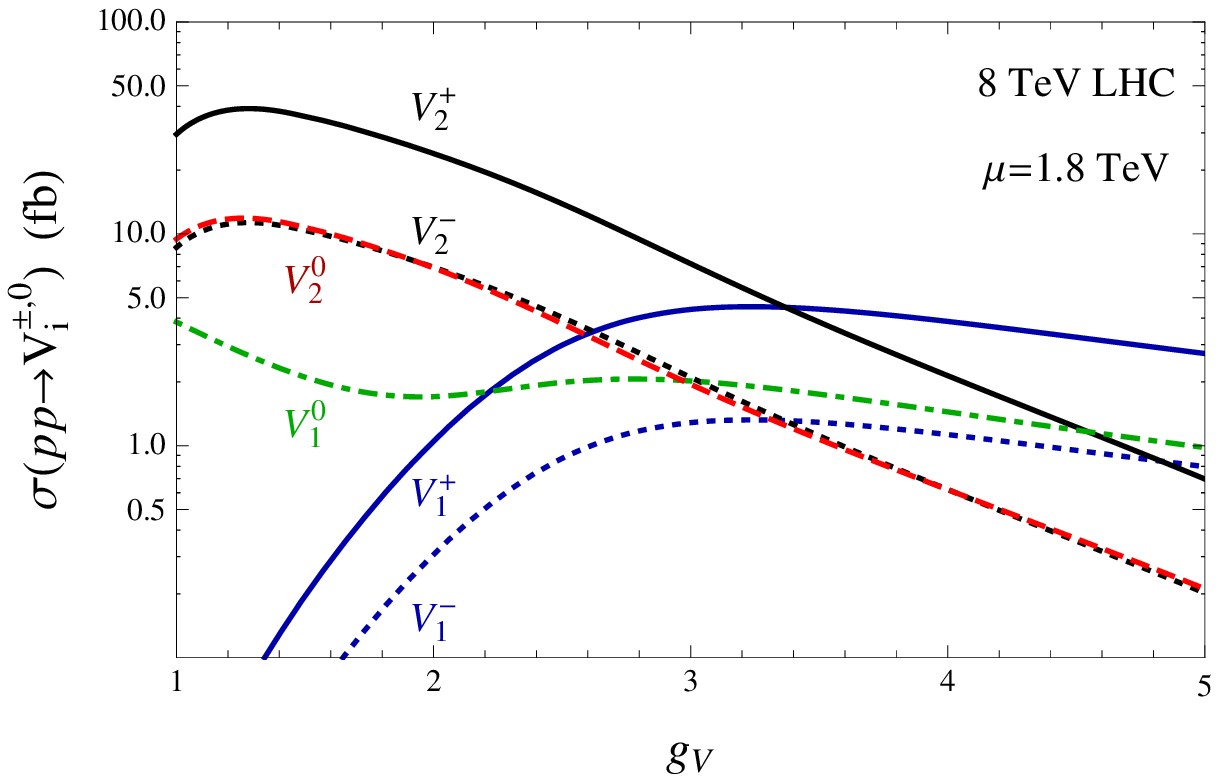}}
\put(230,140){\includegraphics[height=4.5cm]{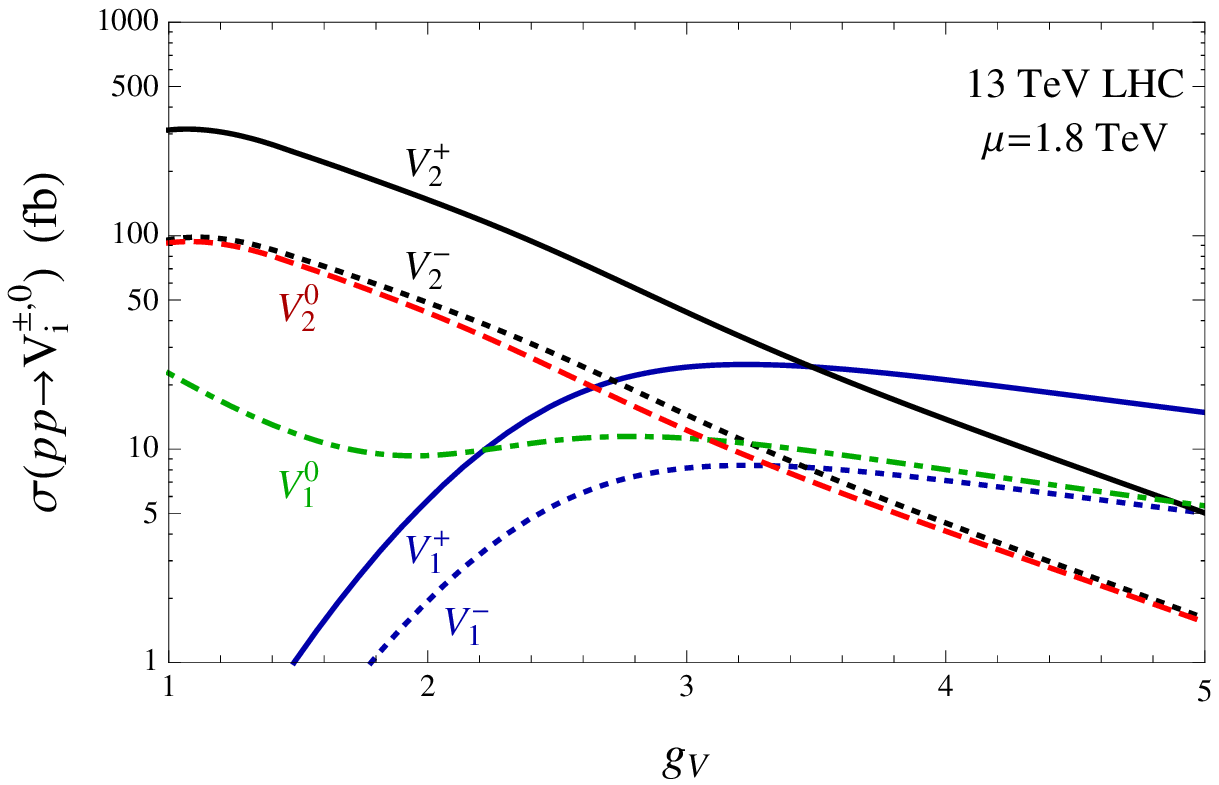}}
\put(120,0){\includegraphics[height=4.5cm]{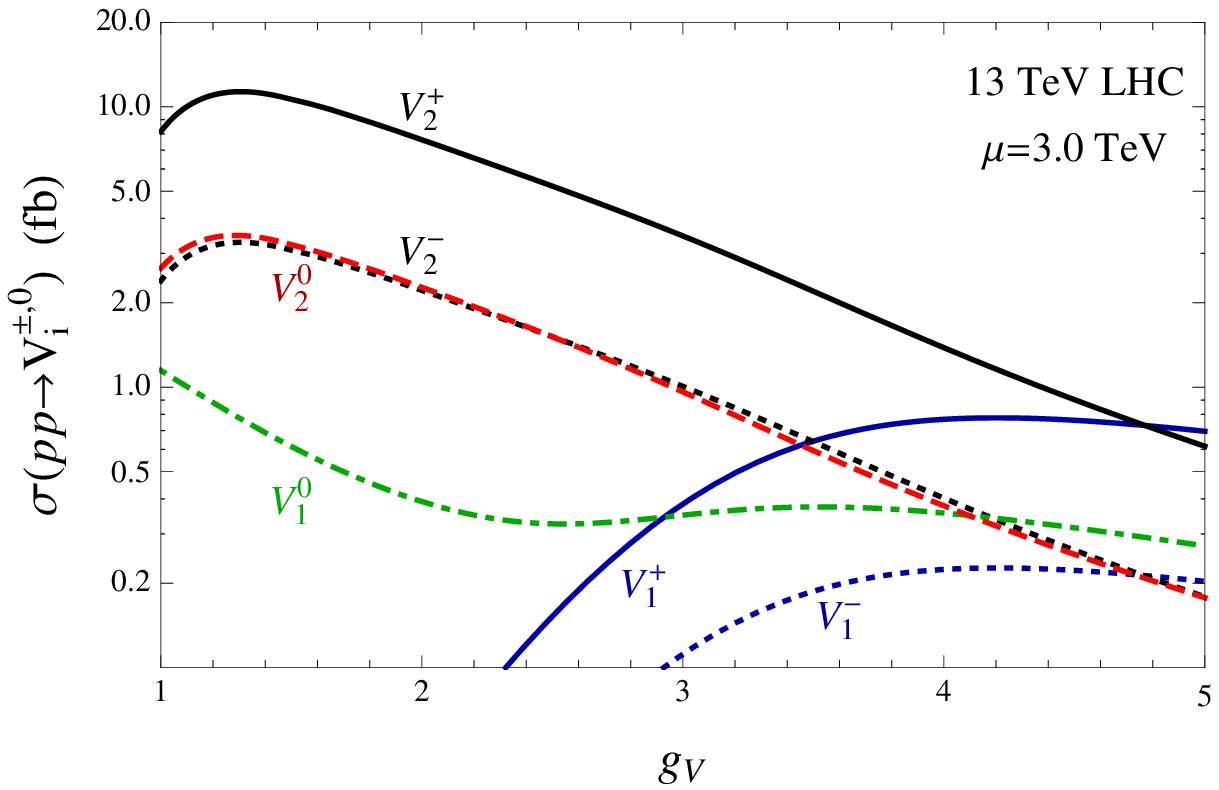}}
\end{picture} 
\caption{Production cross sections in Drell-Yan processes for the heavy vector bosons
in $pp$ collisions at $4+4$ TeV, for $\mu=1.8$ TeV (top left) and at  $6.5+6.5$  TeV 
for $\mu=1.8$ TeV (top right) and $\mu=3.0$ TeV (bottom).}
\label{Fig:DY}
\end{center}
\end{figure}
In Fig.~\ref{Fig:DY}, we show the Drell-Yan (DY) production cross sections for the 
six heavy particles at both $8$ TeV and
$13$ TeV at the LHC. The MSTW 2008 PDFs~\cite{arXiv:0901.0002} have been used in our numerical 
calculations with the renormalization scale chosen to be the resonance mass.
Following the strategy in~\cite{PTTW},
we have checked that the vector-boson-fusion (VBF) production cross 
sections are relatively suppressed, always below
$1$ fb,  throughout the parameter range. In the $pp$ 
collisions of the LHC, the negatively charged particles, which otherwise 
have the same properties as the positively charged ones, have smaller DY 
production cross sections. In the parameter range $ g_{V} \lesssim 3$, where 
$f \ll F$, the production of the $V_{1}^{\pm}$  is strongly 
suppressed relative to the $V_{2}$. The reason is that $V_{1}^{\pm}  
\approx R^{\pm}$ in this range, and is therefore not directly coupled to 
the quarks. At $8$ TeV, the production of the $V_{1}^{0}$ is also 
suppressed relative to the $V_{2}$ due to the smallness of the neutral 
EW coupling relative to the charged EW coupling.

\begin{figure}[h]
\begin{center}
\begin{picture}(440,280)
\put(220,5){\includegraphics[height=4.7cm]{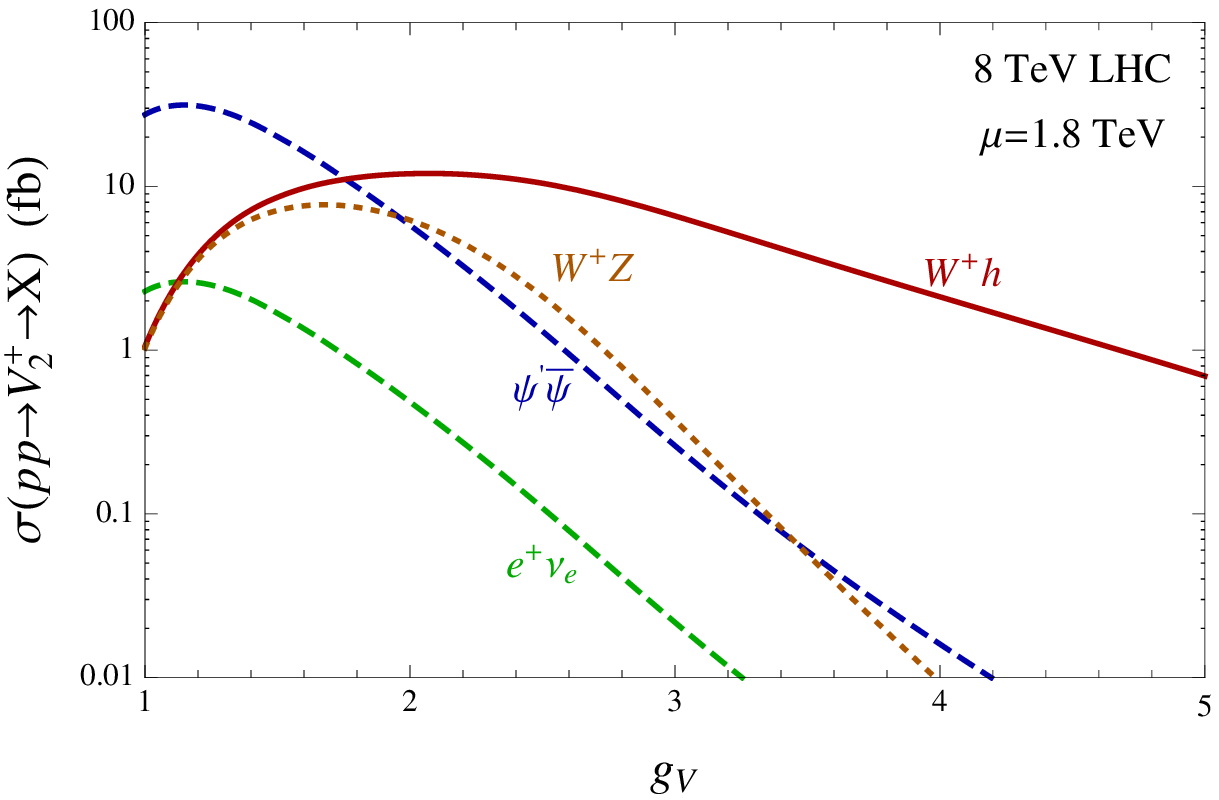}}
\put(0,155){\includegraphics[height=4.7cm]{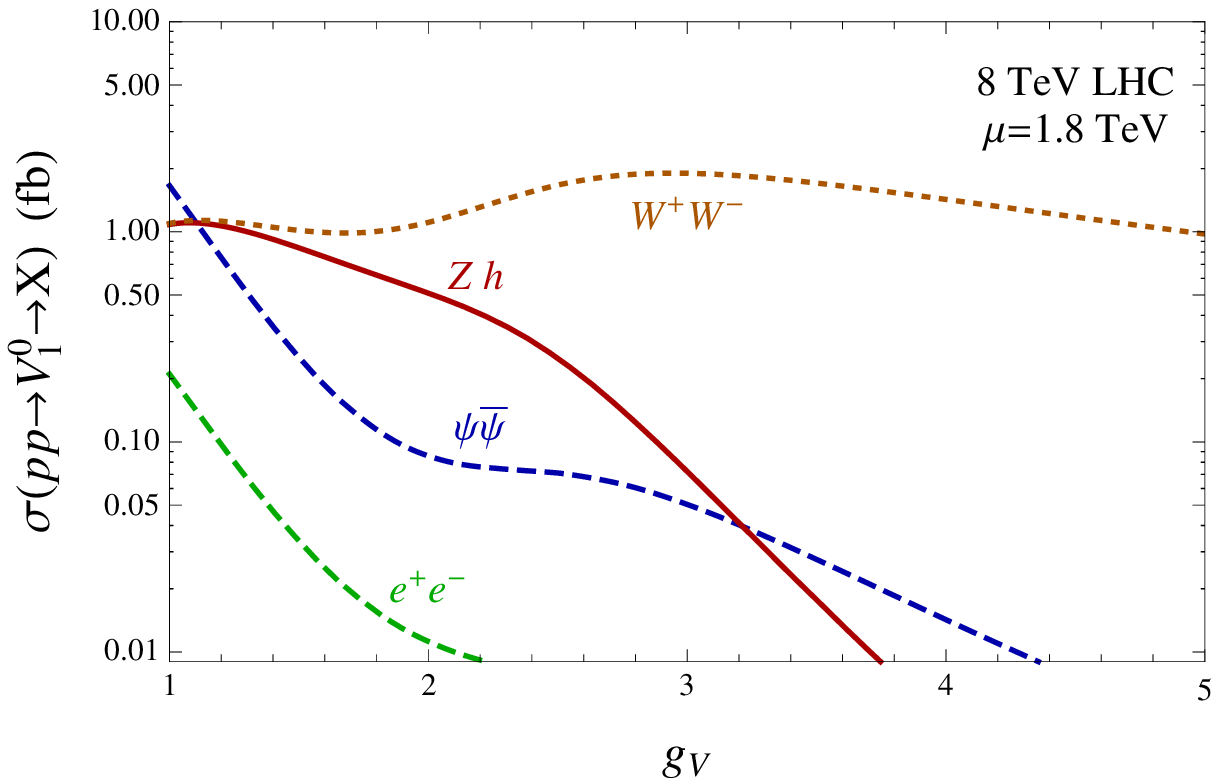}}
\put(0,5){\includegraphics[height=4.7cm]{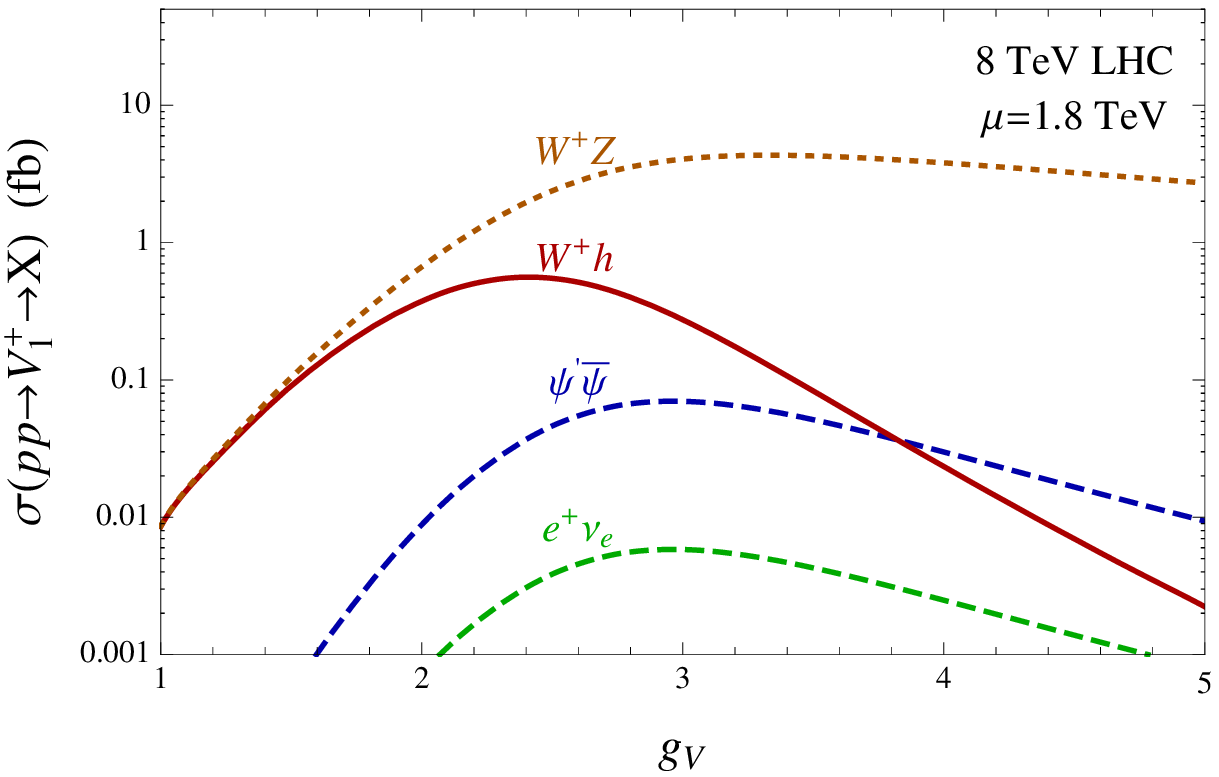}}
\put(220,155){\includegraphics[height=4.7cm]{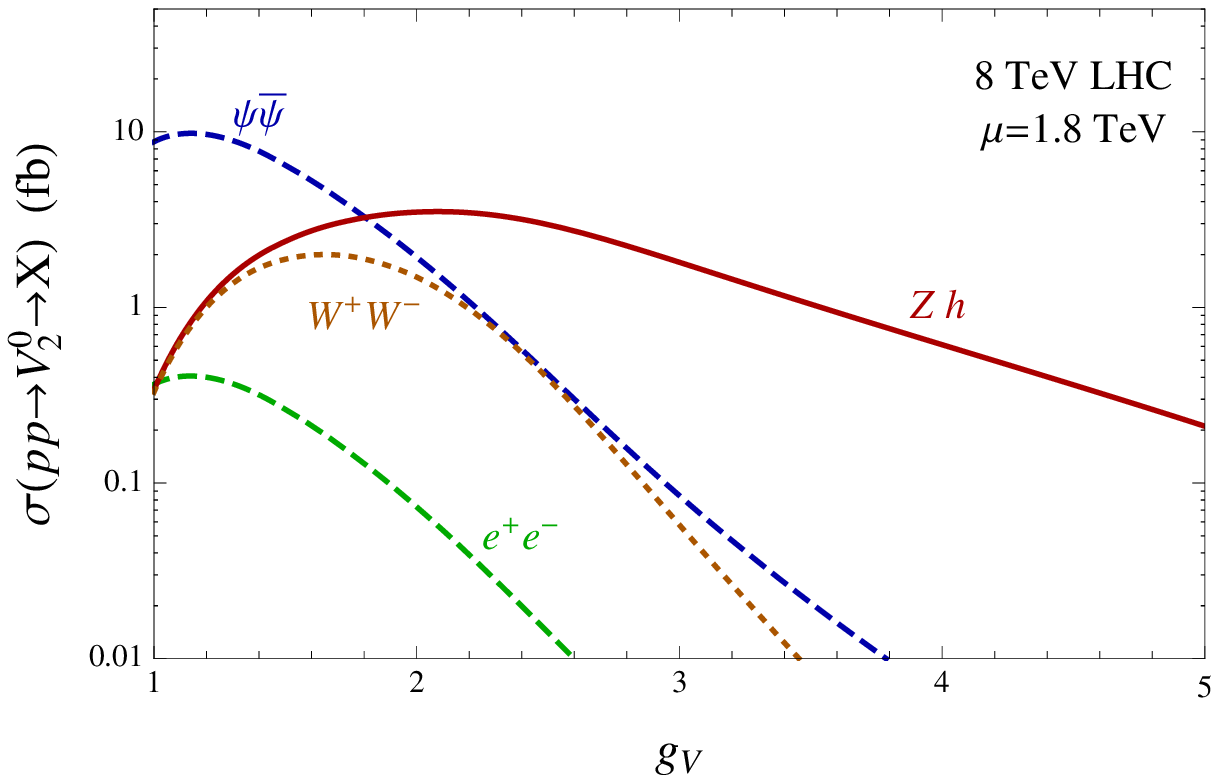}}
\end{picture} 
\caption{Cross section times branching fractions for the heavy particles,
for $\mu=1.8$ TeV for the 8 TeV LHC. The top-left panel shows $V_1^0$, top-right shows $V_2^0$, bottom-left shows $V_1^{+}$
and the bottom-right panel shows $V_2^+$.}
\label{Fig:DYBR}
\end{center}
\end{figure}

In Fig.~\ref{Fig:DYBR}, we show cross sections times branching fractions for the 
case $\mu = 1.8$ TeV,  for $8$ TeV at the LHC.
Results are shown for the $V_2^+$, which has the largest 
      production cross section for much of the $g_V$ range, as well as for the $V_2^0$, $V_1^0$  and $V_1^+$.
 First of all, by requiring that the cross section 
for $pp \rightarrow V_2^{+} \rightarrow e^{+} \nu_{e}$ is bounded by $ 
\sigma \times \mbox{BR} \lsim 0.5$ fb~\cite{ATLASW'},  we deduce the approximate 
bound  $g_V \gsim 2$. Similarly, by requiring that the cross section for 
$pp \rightarrow V_2^{0} \rightarrow e^{+} e^{-}$ is bounded by $ \sigma 
\times \mbox{BR} \leq 0.2$ fb~\cite{ATLASZ'}, we deduce the approximate bound $g_V 
\gsim 1.6$.

We conclude that for $\mu = 1.8$ TeV, the bounds from standard-model 
processes and from direct searches involving new spin-one particles are 
satisfied provided that the coupling $g_V$ is in the range
\beq
2.0 \lsim g_V \lsim 4.2\,,
\label{Eq:bound}
\eeq
with the lower bound being a $95\%$ bound from direct searches, and the 
 $3 \sigma$ upper bound coming from
$pp \rightarrow h \rightarrow WW$. The bounds on the $\hat{S}$ parameter are 
always satisfied at the $3 \sigma$ level.
We note that this allowed coupling range broadly agrees with the range 
emerging from a fit performed in Ref.~\cite{TRW}
in the context of $W^\prime$ models.

In our allowed parameter range, and with $\mu = 1.8$ TeV, the mass of 
the $V_2$ is $1.9 - 2.0$ TeV  while the mass of the $V_1$ is close to 
$1.8$ TeV. Despite the relative heaviness of the $V_2^+$, it is its production and subsequent decay to $WZ$ and $Wh$ that could be 
prominent enough to explain the observed excess in Run 1 of the LHC, at 
least for $g_V$ near the lower end of the above range $g_V\simeq 2$. The cross section 
times branching ratio in these two cases is
$6-7$ fb and $10$ fb respectively (see also the model-independent analysis in~\cite{BHKRT,AGS}).
 The fact that this particle is 
degenerate with the $V_2^0$ and $V_2^-$ yields a modest enhancement in 
these events. By contrast, for small $g_V$, the production of the $V_1^0$ and $V_1^{\pm}$ 
followed by di-boson decay is too small to be observable in Run 1 of the LHC.

\begin{figure}[h]
\begin{center}
\begin{picture}(440,285)
\put(220,5){\includegraphics[height=4.7cm]{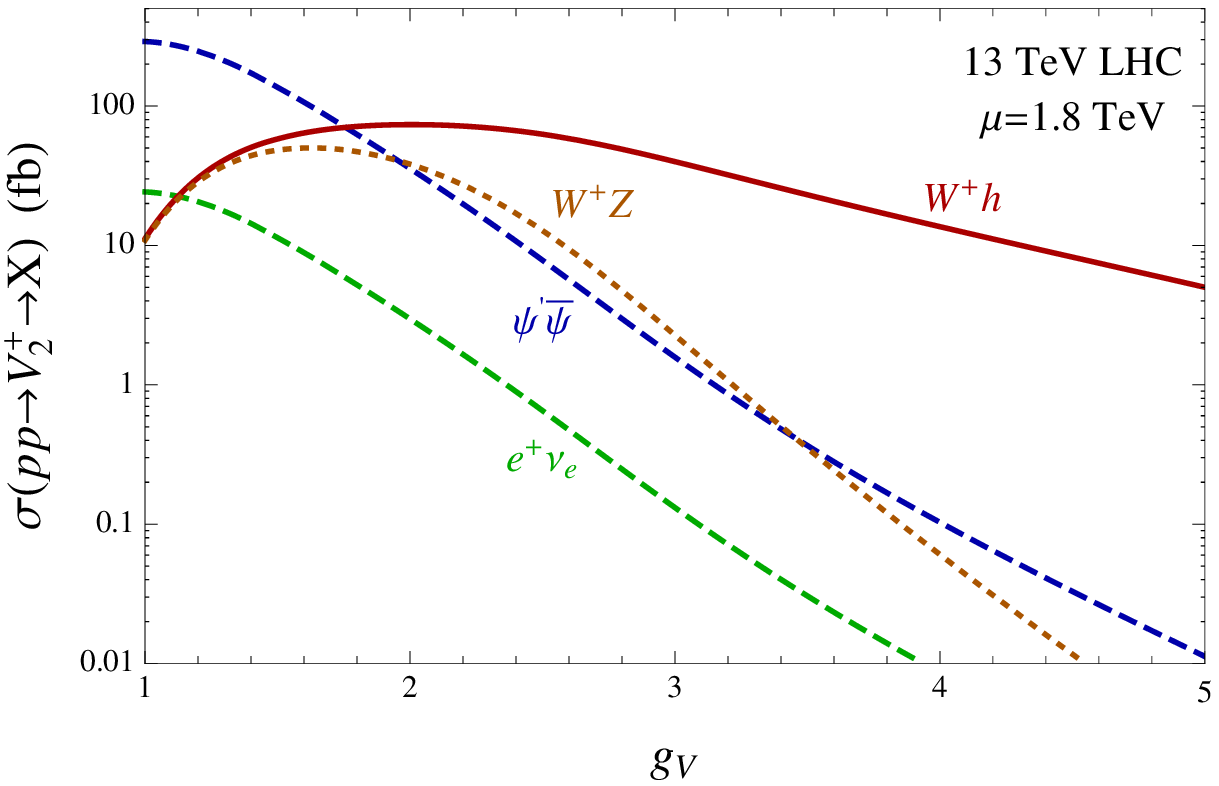}}
\put(0,155){\includegraphics[height=4.7cm]{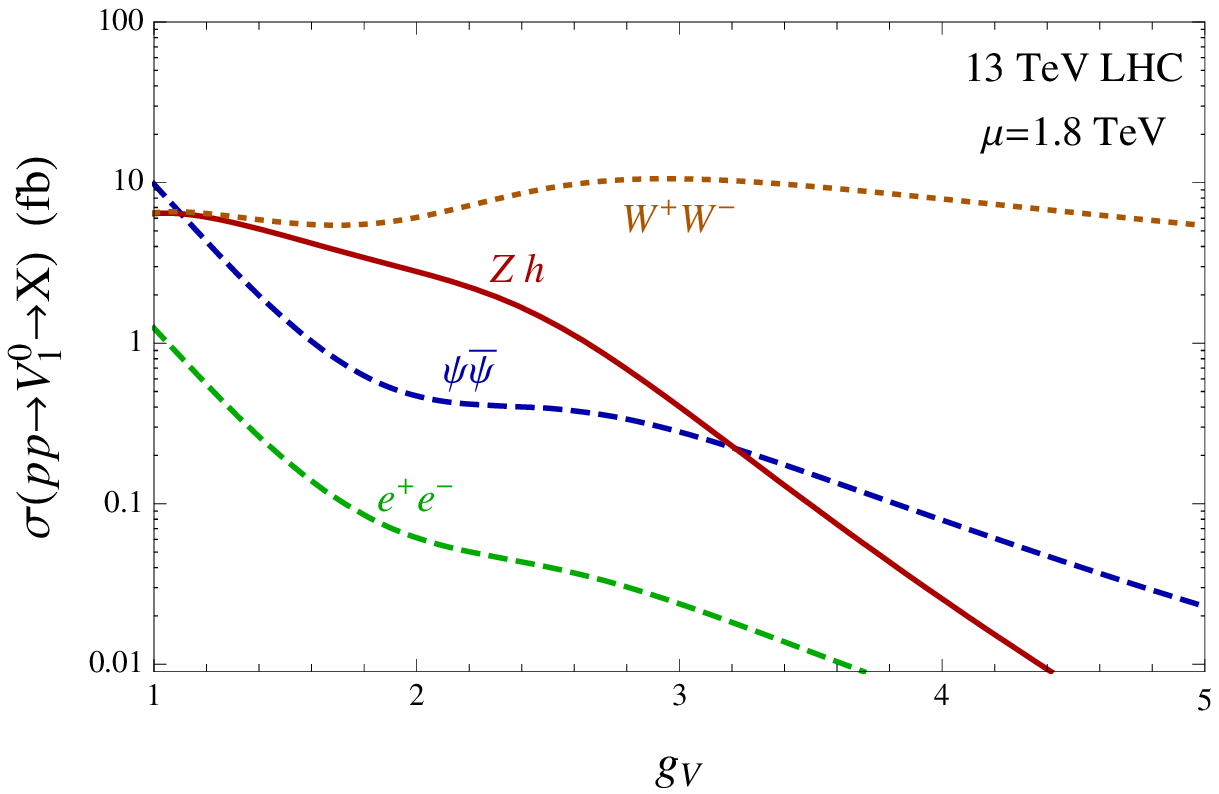}}
\put(0,5){\includegraphics[height=4.7cm]{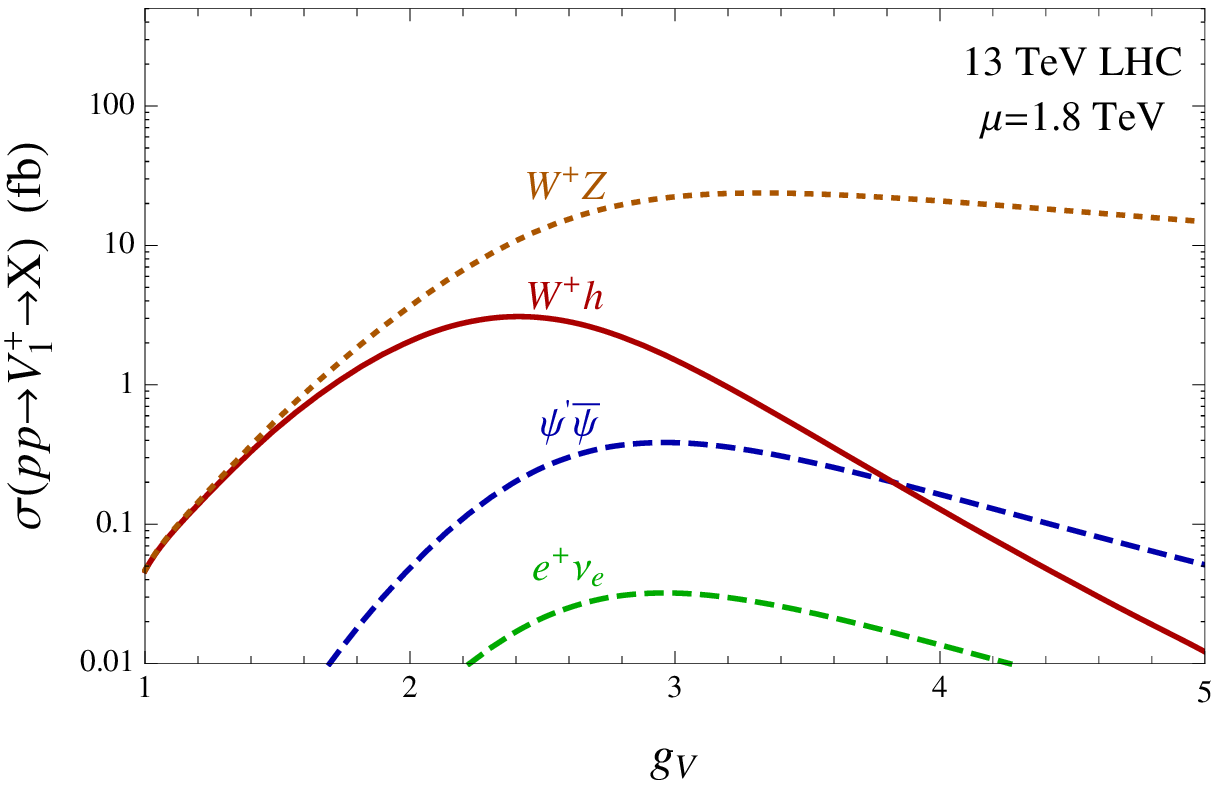}}
\put(220,155){\includegraphics[height=4.7cm]{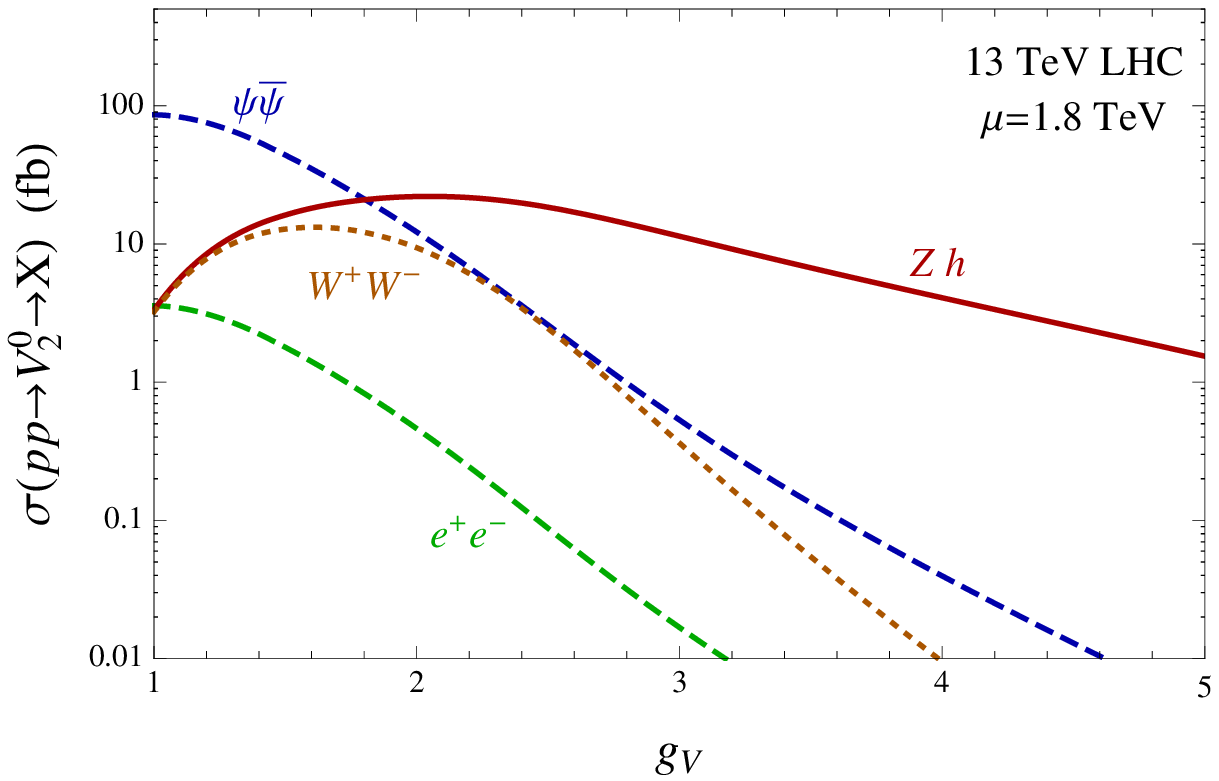}}
\end{picture} 
\caption{Cross section time branching fractions for the heavy particles,
for $\mu=1.8$ TeV at the 
$13$ TeV Run 2 at LHC. Top-left panel shows $V_1^0$, top-right panel shows $V_2^0$, bottom-left panel shows $V_1^{+}$
and bottom-right panel shows $V_2^+$.}
\label{Fig:DYBR13}
\end{center}
\end{figure}

Despite the fact that we have introduced six new vector bosons, for $g_V$ in the lower part of the allowed range only the heaviest, positively charged one is actually observable with the energy and integrated luminosity of Run 1 at the LHC. Its coupling being relatively weak, this particle resembles a heavy $W^{\prime}$ proposed in other contexts~\cite{TRW,BHKRT,Wprime}. The main difference is that we predict the existence of five additional particles. In particular, the three $V_1$ particles have a mass that is $100$ to $200$ GeV smaller, and might have escaped detection because of suppressed couplings to SM fermions and hence suppressed production rates. On the other hand, for larger allowed values of $g_V$, for example $g_V\gtrsim 3$, both $V_1^+\rightarrow W^+Z$ and $V_2^+\rightarrow W^+h$   have cross-section times branching ratios around $4$ fb, and hence might provide a different, more conventional,
 explanation for the  currently observed anomalies.

A distinctive feature of our model is that the new vectors couple to the SM fermions only via mixing with SM gauge bosons, as explained earlier. As a consequence, all the new charged vectors couple only to left-handed fermionic currents. This is atypical for $W^{\prime}$ models. The coupling of the neutral heavy vectors to the SM fermions depends on the value of $g_V$. For the smallest allowed values, the $V_2^0$ couples predominantly to left-handed fermionic currents and the $V_1^0$ couples to a left-right admixture.

In Fig.~\ref{Fig:DYBR13}, we show the product of production cross section times branching ratio
for the new particles with $\mu=1.8$ TeV, but in collisions at $13$ TeV.
For large enough integrated luminosities, all the particles become visible,
including $V_1^+$ and $V_1^0$.
At the time in which we are editing this paper, the LHC collaborations are in the process 
of collecting data at 13 TeV center-of-mass energy: once their combined searches in leptonic as well as di-boson final states
are published, it will be possible to draw exclusion regions in the $(g_V,\mu)$ plane,
and test the viability of the present model.


\section{Discussion}
\label{Sec:discussion}

Motivated by tentative signals from the ATLAS and CMS experiments~\cite{ATLAS,CMS}
for high-mass resonant production
of $WW$, $WZ$, $ZZ$ and $Wh$ pairs, we have developed a simple effective field
theory that extends the standard model to include new vector
resonances. A set of additional $SU(2)_{L^\prime} \times SU(2)_{R^\prime}$ gauge fields is
included, preserving
custodial electroweak symmetry in the new sector. The ingredients and
structure of the EFT are shown in Table~\ref{Fig:4sites}.
The standard model sector of the EFT is described by the usual parameters: gauge
couplings $g$ and $g'$, an electroweak-scale VEV $f$, and fermion Yukawa
couplings. In addition, a left-right symmetric gauge coupling $g_V$ and
a vacuum expectation value $F \gg f$ characterize the new sector.

The structure of the EFT is explored keeping fixed the electroweak gauge-boson masses, the electroweak VEV
$v_W \approx 246$ GeV, and the characteristic scale of new physics. This we take to range upward from the $2$ TeV scale
 suggested by the ATLAS and CMS excesses. For each choice of this scale, the structure of the EFT then depends
on one parameter, which we take to be $g_V$. It can a priori range from electroweak strength up to ${\cal O}(4\pi)$ where the
new gauge bosons are strongly coupled. As $g_V / g$ increases, the ratio $F/f$ must decrease to keep fixed the scale of new physics $\mu = g_{V} F/2$. With this scale taken to be approximately $2$ TeV, it turns out that at least one of the two conditions $f/F\ll 1$ or $g/g_{V} \ll 1$ must be met. Standard model
precision physics, in particular the coupling of the Higgs boson to the $W^{\pm}$ then restricts the value of 
$g_V$ to the range $g_{V}  \lesssim 4.2$. By contrast, direct searches in leptonic channels lead to a lower bound $g_V \gtrsim 2.0$, for $\mu = 1.8$ TeV, in order to  suppress adequately these branching fractions.

In the allowed region of parameter space, the model
 exhibits approximate parity doubling, in the sense that all the new vector
bosons have masses that differ by no more than $100-200$ GeV, and similar widths. 
 Furthermore, for $g_V$ in the lower part of the range shown in Eq.~(\ref{Eq:bound}), the heaviest charged resonance $V_2^+$  
has a production cross section and di-boson branching fraction that could account for both the anomalies reported by ATLAS and
CMS, while lighter new particles, in particular the $V_1^{\pm}$, have a suppressed production cross section, putting 
them below observability at the LHC Run 1. 
By contrast, for $g_V\gtrsim 3$ the process $pp\rightarrow V_2^+\rightarrow Wh$ and 
$pp\rightarrow V_1^+\rightarrow WZ$ might separately be observable.
These features of the model lead to the exciting prospect that future LHC exploration 
at higher energy and luminosity could reveal a rich phenomenology of heavy vector states.

To conclude, the model proposed and studied here is a simple representation of new physics that could arise at higher mass scales,
such as the $2$-TeV scale already accessible in Run 1 of the LHC. There, some possible excesses have been reported,
and future studies in the $13$ TeV Run 2 of the LHC could be even more interesting. The model could also describe new physics 
accessible only at these higher energies. We have used the mass scale $3$ TeV in our model as an example of this.
There, some of the bounds imposed on the parameters of our model would become weaker. Whatever the intrinsic scale of 
our model, it could originate microscopically in a variety of ways. While the coupling $g_V$ is constrained to be relatively
weak in our $2$-TeV model, the underlying dynamics could be weakly coupled or strongly coupled, 
even at experimentally accessible energies.

\vspace{1.0cm}
\begin{acknowledgments}
We would like to thank Bogdan Dobrescu and Kenneth Lane for helpful discussions. 
The work of TA and JI is supported by the U. S. Department of Energy under the contract DE-FG02-92ER-40704. YB is supported by the U. S. Department of Energy under the contract DE-FG-02-95ER40896.
The work of MP is supported in part by the STFC Consolidated Grant ST/J000043/1.

\end{acknowledgments}
\appendix

\section{Relation to Weinberg Sum Rules}
\label{Sec:Weinberg}

The approach described here is in line with that often followed in the 
context of dynamical
electroweak symmetry breaking. This amounts to computing current-current 
correlation functions in the new
strongly-coupled sector responsible for electroweak symmetry breaking, and 
relating this to the
propagators of the electroweak gauge bosons by assuming that the $SU(2)_L \times U(1)_Y$ gauge
group is a subgroup of the symmetry group of the new sector.  This maps 
onto our framework
in the limit $g \ll g_V$, where $V_1 \approx (L+R)/\sqrt{2}$, $V_2 
\approx (L-R)/\sqrt{2}$,
and the new gauge-boson coupling is relatively strong.

A familiar expression for the $S$ parameter in the Weinberg-sum-rule 
context is
\beqs
S&\equiv&4\pi\sum\left(\frac{f_{\r}^2}{M_{\r}^2}-\frac{f_{a_1}^2}{M_{a_1}^2}\right)\,,
\eeqs
where the sum is over all heavy spin-1 bosons, $M_{\r}$ and $M_{a_1}$ are the masses of the vector and axial-vector resonances, respectively,
while $f_{\r}$ and $f_{a_1}$ are their decay constants.

In terms of $\Sigma^+$ (the current-current correlation function of the theory with $g=0=g^{\prime}$) we find
\beqs
\Sigma^+&
\equiv &F^2\,+\,\frac{M_{\r}^2f_{\r}^2}{2(q^2-M_{\r}^2)}\,+\,\frac{M_{a_1}^2f_{a_1}^2}{2(q^2-M_{a_1}^2)}\,,
\eeqs
where one can explicitly show that
\beqs
M_{\r}^2&=&\frac{1}{4}g_{V}^2 F^2\,,~~~
M_{a_1}^2\,=\,\frac{1}{4} g_{V}^2(F^2+2f^2)\,,~~~
f_{\r}^2\,=\,\frac{F^2}{2}\,,~~~
f_{a_1}^2\,=\,\frac{F^4}{2(F^2+2f^2)}\,.
\eeqs

The first and second Weinberg sum rules follow :
\beqs
f_{a_1}^2-f_{\r}^2&=&
-\frac{f^2F^2}{F^2+2f^2}\,=\,-f_{\pi}^2\,,\\
M_{a_1}^2f_{a_1}^2\,-\,M_{\r}^2f_{\r}^2&=&
0\,.
\eeqs
The pion decay constant is defined in the body of the paper
$
\lim_{q\rightarrow 0} \Sigma^+\,=\,
\frac{f^2F^2}{F^2+2f^2}\,=\,f_{\pi}^2
$,
where we identify it with the electroweak scale $f_{\pi}=v_W=246$ GeV.
Notice that this relation is actually exact, and does not rely on taking $g=0$.

Using the above expressions, the $\hat{S}$ parameter becomes:
\beqs
\hat{S}&\equiv&
\frac{1}{4}g^2\left(\frac{f_{\r}^2}{M_{\r}^2}-\frac{f_{a_1}^2}{M_{a_1}^2}\right)
\,=\,\frac{g^2}{g_{V}^2}\,\frac{2f^2(F^2+f^2)}{(F^2+2f^2)^2}\,,
\eeqs
in agreement with Eq.~(\ref{Eq:Sstrong}), appropriate for the limit $g, g^{\prime} \ll
g_V$.
\section{Indirect Bounds from Higgs Physics}
\label{Sec:Higgs}

All the production and decay rates of the Higgs boson are affected by the 
way in which the couplings to the SM fields are suppressed in our model.
We saw that the coupling to fermions is suppressed by the coefficient $c=v_W/f$.
This suppression in turns affects the $hgg$ coupling to the gluons that is responsible for the 
main contribution to the production cross-section of the Higgs particle,
\beqs
\sigma(gg\rightarrow h)&=&c^2\sigma(gg\rightarrow h)_{\rm SM}\,,
\eeqs
as well as the associated production with top quarks:
\beqs
\sigma(pp\rightarrow ht\bar{t})&=&c^2\sigma(pp\rightarrow ht\bar{t})_{\rm SM}\,.
\eeqs
The same coupling $c$ affects the decay rates into SM fermions, as well as gluons (via the top loop):
\beqs
\Gamma(h\rightarrow \psi\bar{\psi})&=&c^2\,\Gamma(h\rightarrow \psi\bar{\psi})_{\rm SM}\,,\\
\Gamma(h\rightarrow gg)&=&c^2\,\Gamma(h\rightarrow gg)_{\rm SM}\,.
\eeqs

The coupling to SM weak gauge bosons is suppressed by the factor $a$.
It affects the production cross section of  processes involving electroweak gauge bosons, such as vector-boson fusion and associated 
production with electroweak gauge bosons $V=W,Z$
\beqs
\sigma(pp\rightarrow hjj)&=&a^2\sigma(pp\rightarrow h j j )_{\rm SM}\,,\\
\sigma(pp\rightarrow Vh)&=&a^2\sigma(pp\rightarrow V h)_{\rm SM}\,.
\eeqs
It also affects the decay to pairs of  electroweak gauge bosons:
\beqs
\Gamma(h\rightarrow WW^{\ast})&=&a^2\,\Gamma(h\rightarrow WW^{\ast})_{\rm SM}\,,
\eeqs
and an analogous formula for $ZZ^{\ast}$.

Finally, the $h\rightarrow \gamma\gamma$ process is additionally affected 
by the Higgs coupling 
to heavy vectors, loops of which contribute to the $h\rightarrow\gamma\gamma$ amplitude.
In the standard model the rate is dominated by the contribution of loops of heavy particles:
\beqs
\Gamma(h\rightarrow\gamma\gamma)_{\rm SM}&=&\frac{G_F\alpha^2m_h^3}{128\sqrt{2}\,\pi^3}\left|A_t(\tau_t)N_cN_fQ^2\,+\,A_W(\tau_W)\right|^2\,,
\eeqs
where $N_c=3$, $N_f=1$ and $Q=2/3$ descend from the contribution of top loops,
with $\tau_t=m_h^2/(4m_t^2)$ and 
\beqs
A_t(\tau_t)&=&\frac{2}{\tau_t^2}\left[\frac{}{}\tau_t+(\tau_t-1){\rm arcsin}^2\sqrt{\tau_t}\right]\,.
\eeqs
The contribution of  $W$ loops is controlled by
\beqs
A_W(\tau_W)&=&-\frac{1}{\tau_W^2}\left[\frac{}{}2\tau_W^2+3\tau_W+3(2\tau_W-1)\,{\rm arcsin}^2\sqrt{\tau_W}\right]\,,
\eeqs
with  $\tau_W=m_h^2/(4M_W^2)$.
Notice that $A_W(\tau_W)\simeq -8.3$, while $A_W(0)=-7$.
In our model, the rate is
\beqs
\Gamma(h\rightarrow\gamma\gamma)&=&\frac{G_F\alpha^2m_h^3}{128\sqrt{2}\,\pi^3}
\left|c\,A_t(\tau_t)N_cN_fQ^2\,+\,a\,A_W(\tau_W)\,+\,(c-a)A_W(0)\right|^2\,,
\eeqs
where besides the modifications of the couplings to the top and photon, we include the loops of 
the two charged heavy vectors, and take the limit $M_{V_{1,2}}\gg m_h$. The couplings satisfy
$a_1+a_2=c-a$. 

The contribution of the di-photon channel to the total width is negligibly small.
In the standard model the BR to $WW^{\ast}$ or $ZZ^{\ast}$ sum up to approximately $25\%$, while the remaining
$75\%$ comes primarily from $bb$, $cc$, $\tau\tau$ and $gg$, meaning that in our model the total width 
scales as
\beqs
\frac{\Gamma}{\Gamma_{\rm SM}}&\simeq&0.75\,c^2\,+\,0.25\,a^2\,.
\eeqs


\end{document}